\documentclass[useAMS,referee,usenatbib]{biom}
\newcommand{\ignore}[1]{}
\usepackage{natbib}
\usepackage{amsfonts,amsmath,amssymb,graphicx,color}
\usepackage{algorithm,comment}
\usepackage[compatible]{algpseudocode}
\def\bbeta{{\boldsymbol \beta}}
\def\bSigma{{\boldsymbol \Sigma}}
\def\R{{\mathbb R}}
\def\Y{{\mathbf Y}}

\def\bx{{\boldsymbol x}}
\def\bA{{\mathbf A}}

\def\S{{\mathcal S}}  

\def\Y{{\mathbf Y}}
\def\E{{\mathbb E}}

\def\diag{{\rm diag}}

\def\var{{\rm Var}}

\def\P{{\mathbb{P}}}

\newcommand{\bX}{\mathbf{X}}
\newcommand{\bZ}{\mathbf{Z}}
\newcommand{\btZ}{\widetilde{\mathbf{Z}}}

\newcommand{\biden}{\mathbf{I}_p}
\newcommand{\tM}{\widetilde{M}}

\DeclareMathOperator*{\argmin}{argmin}

\allowdisplaybreaks

\title[GLMs with  compositional covariates]{Generalized Linear  Models with Linear Constraints for Microbiome Compositional Data} 
\author{Jiarui Lu, Pixu Shi and Hongzhe Li$^*$\email{hongzhe@upenn.edu} \\
	Department of Biostatistics, Epidemiology and Informatics, University of Pennsylvania\\
	Perelman School of Medicine, Philadelphia, PA 19104, U.S.A.
}

\begin{document}
	
	\date{}

	\pagerange{\pageref{firstpage}--\pageref{lastpage}}
	\volume{}
	\pubyear{}
	\artmonth{}

	\doi{}

	\label{firstpage}

\begin{abstract}
Motivated by regression analysis for microbiome compositional data, this paper considers generalized  linear regression analysis with compositional covariates, where a group of linear constraints on regression coefficients are imposed to account for the compositional nature of the data and to achieve subcompositional coherence. A penalized likelihood estimation procedure using a  generalized accelerated proximal gradient  method is developed to efficiently estimate the regression coefficients.  A de-biased procedure is developed  to obtain asymptotically unbiased and normally distributed estimates, which leads to valid confidence intervals of the regression coefficients. Simulations results show the correctness of the coverage probability of the confidence intervals and smaller variances of the estimates when the appropriate linear constraints are imposed. The methods are illustrated by a microbiome study in order to identify bacterial species that are associated with inflammatory bowel disease (IBD) and to predict IBD using fecal microbiome. 

\end{abstract}


\maketitle

\section{Introduction}
Human micorbiome consists of all living microorganisms that are in and on human body. These micro-organisms have been shown to be associated with complex  diseases and to influence our health.  Advanced sequencing technologies such as 16S sequencing and shotgun metagenomic sequencing, provide powerful methods to quantify the relative abundance of bacterial taxa in large samples. Since only the relative abundances are available, the resulting data are compositional with a unit sum constraint.   The compositional nature of the data requires additional care in statistical analysis, including linear regression analysis  \citep{lin2014variable,shi2016,aitchison1984log}.

The main challenges of analyzing compositional data are to account for the unit sum structure and to achieve subcompositional coherence \citep{aitchison1982statistical}, which requires that the same results are obtained regardless of the way the data is normalized based on the whole compositions or only a subcomposition. To explore the association between outcome and the compositional data, \citet{aitchison1984log} proposed a linear log-contrast model to link the outcome and the log of the compositional data. \citet{lin2014variable} further developed  this model and considered  variable selection by  a $\ell_{1}$-penalized estimation procedure.  To achieve  subcompositional coherence, \citet{shi2016} extended  the linear regression model by imposing  a set of linear constraints. The log-contrast model and its extensions are suitable when the outcome variable is continuous and normal distributed. 

In this paper, the  generalized linear regression models (GLMs) with linear constraints in the regression coefficients are proposed  for microbiome compositional data, where a group of linear constraints are imposed to achieve subcompositional coherence. In order to  identify the bacterial taxa that are associated with the outcome, a penalized estimation procedure for the regression coefficients via a $\ell_1$ penalty  is introduced.  To solve the computational problem, a generalized accelerated proximal gradient method is developed, which  extends the standard accelerated proximal gradient method \citep{nesterov2013introductory} to account for linear constraints.  The proposed method can efficiently solve the optimization problem of minimizing the penalized negative log-likelihood subjects to a group of linear constraints.

Previous works on the inference of Lasso for the generalized linear models include \citet{buhlmann2011statistics}, who  provided  properties of the penalized estimates such as bound for $\ell_1$ loss  and oracle inequality. However, the methods cannot be applied directly to the setting with linear constraints. Furthermore, it is known that the $\ell_1$ penalized estimates are biased and do not have a tractable asymptotic distribution. In order to correct such biases, works  have been done for the Lasso estimate, including \citet{zhang2014confidence}, who proposed a low-dimensional projection estimator to correct the bias and \citet{javanmard2014confidence}, who used a quadratic programming method to carry out the task. \citet{van2014asymptotically} considers an extension to generalized linear models.  However, these methods still cannot be directly applied to our problem due to the linear constraints.

 In order to make statistical inference on the regression coefficients,  we propose a bias correction procedure  for GLMs with linear constraints by extending the method of \citet{javanmard2014confidence}. Such a debiased procedure provides  asymptotically unbiased and normal distributed estimates of the regression coefficients, which can be used to construct confidence intervals.   Our simulations results show the correctness of the coverage probability of the confidence intervals and smaller variances of the estimates when the appropriate linear constraints are imposed.

Section \ref{sec: model} develops the  GLMs for compositional data and provides an efficient algorithm to solve the optimization problem. Section \ref{sec: de-biased} provides a de-biased procedure to correct the biases of the penalized estimates and derives the asymptotic distribution of the de-biased estimates. Section \ref{sec: IBD} presents the result of identifying gut bacterial species that are associated with inflammatory bowel disease. Section \ref{sec: simulation} provides the simulation results that  illustrate the correctness of the proposed method. Some discussion and suggestion for future work are provided in Section \ref{sec: discussion}. Proofs of the theorems are included in the Appendix.

\section{GLMs  with Linear Constraints for Microbiome Compositional Data}
\label{sec: model}
\subsection{GLMs with linear constraints}

Consider a microbiome study with  outcome $y_i$ and a $p$ dimensional compositional covariates $\bX_i=(x_{i1}, \cdots, x_{ip})$ with the unit sum  constraint $\sum_{j}x_{ij}=1$ for $i=1, \cdots, n$, where $x_{ij}$ represents the relative abundance of the $j$th taxon for the $i$th samples. 
To account for compositional nature of the covariates,  \citet{lin2014variable} proposed  the linear model with constraint: 
\begin{eqnarray}\label{linear}
y_i = \bZ_i^\top \bbeta + \epsilon_i, \mbox{ subject to } C^\top\bbeta=0
\end{eqnarray}
where $\bZ_i = \{\log(x_{ij})\} \in \R^{n \times p}$ and $C = (1,1, \ldots,1)^\top$.
\citet{shi2016} further developed  this method to  allow multiple linear constraints by specifying the $p\times r$ constraint matrix $C$.  Such  constraints ensure that the regression coefficients  are independent
of an arbitrary scaling of the basis from which a composition is obtained, and remain
unaffected by correctly excluding some or all of the zero components.  This subcompositional coherence property is one of the principals of compositional data analysis \citep{aitchison1982statistical}.

For general outcome, we extend the linear model \eqref{linear} to the generalized linear model with its density function specidied as 
\begin{equation}\label{eq:exp_fam}
\begin{split}
f(y_i|\bbeta, \bZ_i) = h(y_i)\exp \left\{\eta_i y_i-A(\eta_i)\right\},\quad \eta_i = \bZ_i^\top\bbeta\\
\E y_i = \triangledown_{\eta_i} A(\eta_i)\equiv \mu(\bbeta, \bZ_i),\quad \var y_i = \triangledown_{\eta_i}^2 A(\eta_i) \equiv v(\bbeta, \bZ_i) \\
\end{split}
\end{equation}
where $\bbeta = (\beta_{1}, \beta_{2}, \ldots \beta_{p})^\top \in \R^{p}$ and satisfies 
$$C^\top\bbeta=0,$$
 and $\bZ_{i}^\top = (Z_{i1}, Z_{i2}, \ldots, Z_{ip})$.  For simplicity,  we assume the intercept being zero, though our formal justification will allow for an intercept. For binary outcome and logistic regression, we have 
$$A(\eta) = \log(1+e^{\eta}), \mu(\bbeta, \bZ_i) = \dfrac{e^{\bZ_i^\top\bbeta}}{1+e^{\bZ_i^\top\bbeta}}, v(\bbeta, \bZ_i) = \dfrac{e^{\bZ_i^\top\bbeta}}{(1+e^{\bZ_i^\top\bbeta})^2}.$$ 

\subsection{$\ell_1$ penalized estimation with constraints}
The log-likelihood function based on model \eqref{eq:exp_fam} is given by
\begin{align}
\label{eq: loglik}
\ell(\bbeta|\Y,\bZ) &= \sum_{i=1}^n\log h(y_i) + \Y^\top \bZ\bbeta - \sum_{i=1}^nA(\bZ_i^\top\bbeta),
\end{align}
with score function 
and information matrix:
\begin{align*}
\triangledown_{\bbeta}\ell(\bbeta|\Y,\bZ) = (\Y - \boldsymbol{\mu}(\bbeta, \bZ))^\top \bZ, \mbox{  }
\triangledown_{\bbeta}^2\ell(\bbeta|\Y,\bZ) = -\bZ^\top \mathbf{V}(\bbeta,\bZ)\bZ, 
\end{align*}
where $\mathbf{V}(\bbeta,\bZ)=\diag\{v(\bbeta, Z_1),\dots,v(\bbeta, Z_n)\}$. 
The  constraints on $\bbeta$ are given by $C^\top \bbeta =0$, where $C$ is a $p \times r$ matrix.  Without lose of generality, the columns of $C$ are assumed to be orthonormal. Define $P_C = CC^\top$, $\btZ=\bZ(\biden-P_C)$ and $\widetilde{Z}_i = (\biden-P_C)\bZ_i$, then under the constraints of $C^\top\bbeta=0$, all the $\bZ$ and $Z_i$ can be replaced by $\btZ$ and $\widetilde{Z}_i$ because $\bZ\bbeta=\btZ\bbeta$.
 
In high-dimensional settings, $\bbeta$ is assumed to be $s$-sparse, where $s = \#\{i: \bbeta_i \neq 0\}$ and $s = o(\sqrt{n} / \log p)$.  The $\ell_1$ penalized estmates of $\bbeta$ is given as the  solution to the following problem:
\begin{equation}\label{eq:L1}
\hat{\bbeta}^n = \argmin_{\bbeta} \left\{-\dfrac{1}{n}[\Y^\top \btZ\bbeta - \sum_{i=1}^nA(\widetilde{\bZ}_i^\top\bbeta)]+\lambda||\bbeta||_1\right\} \mbox{ subject to } C^\top\bbeta=0,
\end{equation}
where $\lambda$ is a tuning parameter. 

\subsection{Generalized accelerated proximal gradient method}
Due to the linear constraints in the optimization problem \eqref{eq:L1}, the standard coordinate descent algorithm cannot be applied directly. We develop  a  generalized accelerated proximal gradient algorithm. Specifically,  define $g,h$ as following
\[
g(\bbeta) = -\dfrac{1}{n}[Y^\top \btZ\bbeta - \sum_{i=1}^nA(\widetilde{Z}_i^\top\bbeta)], \quad h(\bbeta) = \lambda||\bbeta||_{1}
\]
so the optimization problem \eqref{eq:L1} becomes 
\[
\hat{\bbeta}^n = \argmin_{\bbeta} \left\{ g(\bbeta)+ h(\bbeta)\right\} \mbox{ subject to } C^\top\bbeta=0.
\]
Since  $g$ is convex and differentiable and $h$ is convex, the standard accelerated proximal gradient method \citep{nesterov2013introductory} is given by the following iterations:
\begin{align*}
& \bbeta^{(k)}  = \textbf{prox}_{t_{k}h} \left(y^{(k-1)} - t_{k} \nabla g(y^{(k-1)}) \right), \\
& y^{(k)}  = \bbeta^{(k)} + \frac{k-1}{k+r-1} (\bbeta^{(k)} - \bbeta^{(k-1)}), 
\end{align*}
where $t_{k}$ is the step size in the $k$-th iteration and $r$ is a friction parameter. The proximal mapping of a convex function $h$, which is the key ingredient of this algorithm, is defined as:
\[
\textbf{prox}_{h} (x) = \argmin_{u} \left\{  h(u) + \frac{1}{2}||x - u||^{2}_{2}\right\}.
\]

We generalize this method to  handle the linear constraints. Denote $S_{C} = \{\bbeta \in \R^{p} \  |  \ C^\top \bbeta =0\}$,  a linear subspace of $\R ^p$. The generalized accelerated proximal gradient method becomes 
\begin{align}
& \bbeta^{(k)}  = \argmin_{\bbeta \in S_{C}} \left\{ \lambda t_{k} \|\bbeta\|_{1}  + \frac{1}{2} || y^{(k-1)} - t_{k} \nabla g(y^{(k-1)}) -\bbeta ||^{2}_{2}\right\}  \label{Prox},\\ 
& y^{(k)}  = \bbeta^{(k)} + \frac{k-1}{k+r-1} (\bbeta^{(k)} - \bbeta^{(k-1)}).
\end{align}
The minimization of \eqref{Prox} can be solved by soft thresholding and projection:
\[
\bbeta^{(k)}  =\Pi_{S_{C}} \left( S_{t_{k} \lambda} \left(y^{(k-1)} - t_{k} \nabla g(y^{(k-1)})\right) \right),
\] 
where linear operator $\Pi_{S_{C}} (u)$ projects $u$ onto space $S_{C}$.  
Since $C^\top$ is a matrix and can be  regarded   as a linear mapping from $R^{p} \mapsto R^{r}$, we have  $S_{C} = \ker (C^\top)$. 
Denote $u_p  = \Pi_{S_{C}} (u)$, we have:
\[
C^\top (u - u_p) = C^\top u
\]
So $u - u_p$ is given by least square estimates: $ u - u_p = (CC^\top)^{\dagger}CC^\top u$, where $A^\dagger$ is the Moore-Penrose pseudo inverse of a matrix $A$. Hence,
\[
\Pi_{S_{C}} (u) = u -  (CC^\top)^{\dagger}CC^\top u.
\]

The step size $t_{k}$ can be fixed or chosen by line search. The procedure of line search consists of the following iterations:  
we start with a initial $t = t_{k-1}$ and repeat $t = 0.5 t$ until the following inequality holds:
\[
g(y - tG_{t}(y)) \leq g(y) - t \nabla g(y)^{\top} G_{t}(y) + \frac{t}{2}\|G_{t}(y)\|_{2}^{2}
\] 
where $y = y^{(k-1)}$. For the friction parameter $r$, \citet{su2014differential} suggested that $r > 4.5$ will lead to fast convergence rate and is set to 10.   

\section{De-biased Estimator and its Asymptotic Distribution}
\label{sec: de-biased}
We collect here  the notations  used in the rest of the paper.  For a vector $\bx$, $\|\bx\|_{p}$ is the standard $\ell_{p}$-norm. For a  matrix $\bA_{m \times n}$, $\|\bA\|_{p}$ is the $\ell_{p}$ operator norm defined as $\|\bA\|_{p} = \sup_{\|x\|_{p}=1}\|\bA \bx\|_{p}$. In particular, $\|\bA\|_{\infty} = \max_{1\leq i \leq m} \sum_{j=1}^{n} |a_{ij}|$ and $|\bA|_{\infty}$ is defined as $|\bA|_{\infty} = \max_{i,j}|a_{ij}|$. For square matrix $\bA$, denote $\sigma_{\textrm{max}}(\bA)(\sigma_{\textrm{min}}(A))$ is the largest (smallest) non-zero eigenvalue of $\bA$.   
 
\subsection{A de-biased Estimator}
Since $\widehat{\bbeta}^n$ in equation \eqref{eq:L1} is a biased estimator for $\bbeta$  due to $\ell_1$ penalization,  we propose the following de-biased procedure, detailed as Algorithm \ref{alg: de-biased}, to obtain asymptotically unbiased estimates of $\bbeta$. 

\begin{algorithm}[h]
	\caption{Constructing a de-biased estimator}\label{alg: de-biased}
	\textbf{Input:} $\Y$, $\bZ$, $\widehat{\bbeta}^n$, and $\gamma$. 
	\textbf{Output:} $\widehat{\bbeta}^u$
\begin{algorithmic}[1]
\State
		Let $\widehat{\bbeta}^n$ be the regularized estimator from optimization problem~\eqref{eq:L1}.
\State
		Set $\btZ=\bZ(\biden-P_C)$, $\widehat{\bSigma} =  (\btZ^\top \mathbf{V}(\hat{\bbeta}^n,\btZ)\btZ)/n$.
\State
		\textbf{for} $i=1,2,\dots,p$ \textbf{do}
		\State
		Let $m_i$ be a solution of the convex program:
		\begin{equation}\label{eq:opt}
		\begin{split}
		\mbox{minimize }& m^\top\widehat{\bSigma}m\\
		\mbox{subject to }& ||\widehat{\bSigma}m-(\biden-P_C )e_i||_{\infty}\leq \gamma. \\
		\end{split}
		\end{equation}
		where $e_{i} \in \R^{p}$ is the vector with one at the $i$-th position and zero everywhere else.
		\State
		Set $M=(m_1,\dots,m_p)^\top$, set
		\begin{equation}\label{eq:tildeM}
		\tM=(\biden-P_C)M.
		\end{equation}
		\State
		Define the estimator $\widehat{\bbeta}^u$ as follows:
		\begin{equation}\label{eq:algorithm}
		\widehat{\bbeta}^u=\widehat{\bbeta}^n+\dfrac{1}{n}\tM\btZ^\top(\Y-\boldsymbol{\mu}(\widehat{\bbeta}^n, \btZ)).
		\end{equation}
\end{algorithmic}
\end{algorithm}

From the construction of $\hat{\bbeta}^u$, it is easy to check that $\hat{\bbeta}^u$ still satisfies $C^\top\hat{\bbeta}^u = 0$.  To provide insights into this algorithm, 
using the  mean value theorem, there exists $\bbeta_{i}^{0}$ such that
\begin{equation*}
\mu(\hat{\bbeta}^n,\bZ_i) - \mu(\bbeta,\bZ_i) = v(\bbeta_{i}^0,\bZ_i)\bZ_i^\top (\hat{\bbeta}^n-\bbeta), \ i=1,2, \ldots, n. 
\end{equation*}
 Define $\widehat{\bSigma}^{0} = (\btZ^\top \mathbf{V}(\bbeta^{0},\btZ)\btZ)/n$, where $\mathbf{V}(\bbeta^{0},\btZ) = \diag\{v(\bbeta_{1}^{0}, Z_1),\dots,v(\bbeta_{n}^{0}, Z_n)\}$, we   have
\begin{align*}
\sqrt{n}\left( \hat{\bbeta}^u - \bbeta \right)
\ignore{&= \sqrt{n}\left(\hat{\bbeta}^n-\bbeta+\dfrac{1}{n}\tM\btZ^\top(Y-\boldsymbol{\mu}(\bbeta,\btZ)) - \dfrac{1}{n}\tM\btZ^\top(\boldsymbol{\mu}(\hat{\bbeta}^n,\btZ)-\boldsymbol{\mu}(\bbeta,\btZ)) \right)\\
& = \sqrt{n}\left((\biden - \tM\widehat{\bSigma^{0}})(\hat{\bbeta}^n-\bbeta)+\dfrac{1}{n}\tM\btZ^\top(Y-\boldsymbol{\mu}(\bbeta,\btZ)) \right)\\}
&= \sqrt{n}[(\biden-P_C) - \tM\widehat{\bSigma}^{0}](\hat{\bbeta}^n-\bbeta)+\dfrac{1}{\sqrt{n}}\tM\btZ^\top(Y-\boldsymbol{\mu}(\bbeta,\btZ))  \tag{$*$} \label{difference} \\
& \equiv \Delta +R.  
\end{align*}
Define $\bSigma =  (\btZ^\top \mathbf{V}(\bbeta,\btZ)\btZ)/n$ and $\bSigma_{\bbeta} = \E \bSigma = \E(v(\bbeta,\widetilde{Z}_1)\widetilde{Z}_1\widetilde{Z}_1^\top)$, and suppose $\bSigma_{\bbeta} = V_{\bbeta}\Lambda_{\bbeta}V_{\bbeta}^\top$ is the eigenvalue decomposition of $\bSigma_{\bbeta}$. Since  $(V_{\bbeta},C)$ is full rank and orthonormal, we have 
$$\bSigma_{\bbeta}=(V_{\bbeta},C)\left(\begin{array}{cc}
\Lambda & 0 \\
0 & 0
\end{array}\right)(V_{\bbeta},C)^\top, \mbox{  }
\Omega_{\bbeta}=(V_{\bbeta},C)\left(\begin{array}{cc}
\Lambda_{\bbeta}^{-1} & 0 \\
0 & 0
\end{array}\right)(V_{\bbeta},C)^\top, 
$$
which implies 
$$\bSigma_{\bbeta}\Omega_{\bbeta}=(V_{\bbeta},C)\left(\begin{array}{cc}
\mathbf{I}_{p-r} & 0 \\
0 & 0
\end{array}\right)(V_{\bbeta},C)^\top=V_{\bbeta}V_{\bbeta}^\top=\biden-P_C.$$
So Step 4 of Algorithm \ref{alg: de-biased}  approximates  $\Omega_{\bbeta}$ by rows.

\subsection{Asymptotic distribution}
In order to derive the asymptotic distribution of the de-biased estimator $\hat{\bbeta}^u$, several   regularity conditions are required. 
\begin{enumerate}
	\item[C1.] \label{bound on I-Pc} $\|\biden - P_{C}\|_{\infty} \leq k_{0}$ for a constant $k_{0}$ that is free of $p$.
	\item[C2.] \label{bound on diagonal} The diagonal elements of $\biden - P_{C}$ are greater than zero. 
\end{enumerate}
 Conditions  C1 and C2 have been used in \citet{shi2016} and naturally hold in our setting as well. 
In addition, define $\btZ^{*} = D\btZ$, where $D\in \widetilde{D}_{ab}$ is defined as:
$$ \widetilde{D}_{ab}= \{D \in \R^{n \times n}: \diag(d_{1}, d_{2}, \ldots, d_{n}), a \leq d_{i} \leq b , \, 0 < a < b \}.$$ 
For any matrix $A \in \R^{n \times m}$, the upper and lower restricted isometry property (RIP) constant of order $k$, $\delta_{k}^{+}(A)$ and $\delta_{k}^{-}(A)$,  are defined as:
\begin{align*}
\delta_{k}^{+}(A) = \sup \left(\dfrac{\|A\alpha\|_{2}^{2}}{\|\alpha\|_{2}^{2}}: \ \alpha \in \R^{m} \ \textrm{is $k$-sparse vector}\right), \\
\delta_{k}^{-}(A) = \inf \left(\dfrac{\|A\alpha\|_{2}^{2}}{\|\alpha\|_{2}^{2}}: \ \alpha \in \R^{m} \ \textrm{is $k$-sparse vector}\right).
\end{align*}
\ignore{and the restricted orthogonal constant (ROC) of order $k_{1}$ and $k_{2}$ is 
\begin{align*}
\theta_{k_{1}, k_{2}}(A) = \sup \left(\dfrac{|\langle A\alpha_{1}, A\alpha_{2} \rangle|}{\|\alpha_{1}\|_{2}\|\alpha_{2}\|_{2}}: \ \alpha_{1}  \ \textrm{is $k_{1}$-sparse vector}, \alpha_{2}  \ \textrm{is $k_{2}$-sparse vector}  \right)
\end{align*}}
We assume  the following RIP condition:
\begin{enumerate}[3.]
	\item[C3.] $\inf_{\widetilde{D}_{01}} \left((3\tau-1)\delta_{2s}^{-}(\btZ^{*}/\sqrt{n}) -(\tau+1)\delta_{2s}^{+}(\btZ^{*}/\sqrt{n}) \right)\geq 4\tau\phi_{0}$ for some constant $\phi_{0}$.   
\end{enumerate}
Condition C3 is slightly stronger than the one used for linear regression, which here we require the inequality holds uniformly over a set of matrices.  The following theorem quantifies the difference between $\hat{\bbeta}^{n}$ and $\bbeta$ in $\ell_1$ norm.
\begin{theorem}
	\label{thm: consistency}
	Let $\hat{\bbeta}^{n}$ be the solution for (\ref{eq:L1}), where $\bbeta$ is $s$-sparse. If Conditions C1-C3 hold,  and the tuning parameter $\lambda = \tau \tilde{c}\sqrt{(\log p)/n}$, then 
\[
\P	\left( \|\hat{\bbeta}^{n} - \bbeta\| _1  \geq \dfrac{s\lambda(k_0+1/\tau)}{\phi_{0}} \right)  \leq  2p^{-c'}
\]
where
$c^{\prime} =  \dfrac{\tilde{c}^{2}}{2K^2}-1$ and $K =\max_{i} \sqrt{(\btZ^\top\btZ/n)_{i,i}}$.
\end{theorem}
\ignore{Furthermore, define the sub-exponential norm of a random variable $X$, denoted by $\|X\|_{\psi_{1}}$, as:
\[
\|X\|_{\psi_{1}} = \sup_{p \geq 1}p^{-1}(\E|X|^P)^{1/p}
\]
}

In order to establish the asymptotic distribution of the de-biased estimates, additional conditions are required:
\begin{enumerate}
\item[C4.] There exist uniform constants $C_{\text{min}}$ and $C_{\text{max}}$ such that $0< C_{\text{min}} \leq \sigma_{\textrm{min}}(\bSigma_{\bbeta}) \leq \sigma_{\textrm{max}}(\bSigma_{\bbeta})\leq C_{\text{max}} < \infty$. 
\item[C5] $| \Omega_{\bbeta}\Theta|_{\infty} < \infty$. 
\item[C6] The variance function $v(\bbeta, \bZ_i)$ satisfies Lipschitz condition with constant $C$; 
 \item[C7] There exists a uniform constant $\kappa>0$ such that  $\|\Omega^{1/2} \widetilde{Z}_{k}\|_{\psi_{2}} \leq \kappa $ for all $k=1,\ldots, n$. 
  \end{enumerate}
In Condition C7,  the sub-Gaussian norm of a random vector $Z \in \R^{n}$ is defined  as
 \[
 \|Z\|_{\psi_{2}} = \sup\left( \|Z^\top x\|_{\psi_{2}}: \ x\in \R^{n} \, \text{and} \, \|x\|_{2}=1\right),
 \]
 and 
the sub-Gaussian norm for a random variable $X$, is defined as
\[
\|X\|_{\psi_{2}} = \sup_{q \geq 1}q^{-1/2}(\E|X|^q)^{1/q}.
\] 

 Conditions C4 and C7 are bounded eigenvalue assumption and bounded  sub-Gaussian norm that  are widely used in the literature of inference with respect to Lasso type estimator \citep{shi2016, javanmard2014confidence}. Condition C5 eliminates extreme situations on $| \Omega_{\bbeta}\Theta|_{\infty}$, which actually can be relaxed to hold in probability. For logistic regression, similar conditions are used in \citet{ning2017general}.  Condition C6 is a Lipschitz condition on the variance function, which holds for many of the GLMs  including logistic regression. 

The following Lemma shows that if the tuning parameter $\gamma$ in the optimization problem \eqref{eq:opt} is chosen to be $c \sqrt{(\log p) / n}$, then $\Omega_{\bbeta}$ is in the feasible set with a large probability. 
\begin{lemma}
	\label{thm: feasible set}
	 Denote $\Theta=   \E \widetilde{Z}_{1}\widetilde{Z}_{1}^\top $. Suppose Conditions C1-C7 hold,   then for any constant $c >0$, the following inequality holds:
	\[
	\P \left(  |\bm{\Omega}_{\bbeta}\widehat{\bSigma} -(\biden-P_C)|_{\infty}  \geq 
	c \sqrt{(\log p) / n} \right)   \leq   2p^{-c_{1}^{''}}+2p^{-c_{2}^{''}}
	\]
	where $c_{1}^{''} = c^{2}C_{\text{min}}/ (24e^2 C_{\text{max}} \kappa^4)-2$ and $c_{2}^{''} = \hat{c}^{2}/2K^2-1$, with $\hat{c} = c \phi_{0}/C|\bm{\Omega}_{\bbeta}\Theta|_{\infty}s(k_0 \tau +1)$ 
	and  $K =\max_{i} \sqrt{(\btZ^\top\btZ/n)_{i,i}}$.
	
\end{lemma}

The following Theorem provides   the bound on $ \|\Delta\|_{\infty}$ and also the asymptotic distribution of the de-biased estimates. 
\begin{theorem}
	\label{thm:bound on Delta}
	For $\Delta= \sqrt{n} [(\biden-P_C) - \tM\widehat{\bSigma}^{0}](\hat{\bbeta}^n-\bbeta)$, if conditions C1-C7 hold, then for $n$ large enough, 
	\[
	\sqrt{n}(\hat{\bbeta}^u - \bbeta) = R + \Delta,
	\]
	where $R|\bZ \rightarrow N(0,\tM \widehat{\bSigma} \tM^\top)$ in distribution and $ \|\Delta\|_{\infty}$ converge to $0$ as $n,p \rightarrow \infty$,  i.e., 
	\[
	\P\left(  \|\Delta\|_{\infty} > \dfrac{c\tilde{c}k_{0}(k_0\tau +1)}{\phi_0} \cdot \dfrac{s\log p}{\sqrt{n}}  \right) \leq 2p^{-c'} +2p^{-c_{1}^{''}}+6p^{-c_{2}^{''}}
	\]
	for some constants $c^\prime$, $c_1^{\prime\prime}$ and $c_1^{\prime\prime}$ defined in Theorem \ref{thm: consistency} and Lemma \ref{thm: feasible set}.

	\ignore{where $K =\max_{i} \sqrt{(\btZ^\top\btZ/n)_{i,i}}$ and
	\[
	c^{\prime} =  \dfrac{\tilde{c}^{2}}{2K^2}-1, c_{1}^{''} = \dfrac{c^{2}C_{\text{min}}}{ 24e^2 C_{\text{max}} \kappa^4}-2, c_{2}^{''} = \dfrac{\hat{c}^{2}}{2K^2}-1,
	\]
	with $\hat{c} = c \phi_{0}/C|\bm{\Omega}_{\bbeta}\Theta|_{\infty}s(k_0 \tau +1)$.}
\end{theorem}

This theorem allows us to obtain the confidence intervals for the regression coefficients, which can be used to further select the variables based on their statistical significance. 
\ignore{To sum up, we have the following conclusion on the asymptotic distribution of the de-biased estimator:
\begin{theorem}
\label{thm: summary}
Supposed Conditions 1-7 hold, with $\hat{\bbeta}^{n}$ be the solution for \eqref{eq:L1} and $\hat{\bbeta}^{u}$ defined in \eqref{eq:algorithm}, we have:
\[
\sqrt{n}(\hat{\bbeta}^u - \bbeta) = R + \Delta,
\]
where $R|\bZ \rightarrow N(0,\tM \widehat{\bSigma} \tM^\top)$ in distribution and $ \|\Delta\|_{\infty}$ converge to $0$ as $n,p \rightarrow \infty$.
\end{theorem}
}

\subsection{Selections of tuning parameters}
The tuning parameter $\lambda$ in \eqref{eq:L1} can be  selected using extended Bayesian information criterion (EBIC) \citep{chen2008extended}, which is an extension of the standard BIC in high dimensional cases. Specifically,  denote $\hat{\bbeta}^{n}_{\lambda}$  the solution of \eqref{eq:L1} using $\lambda$ as the tuning parameter, the EBIC  is defined as 
\[
\text{EBIC}(\hat{\bbeta}^{n}_{\lambda}) = -2\ell(\hat{\bbeta}^{n}_{\lambda}|y,\bZ) + \nu(\hat{\bbeta}^{n}_{\lambda}) \log n + 2\nu(\hat{\bbeta}^{n}_{\lambda})\xi \log p,
\]
where $\nu(s)$ is the number of none zero components of $s$. The choice of $\xi$ is to solve for $p = n^\delta$ and set $\xi = 1- 1/ (2 \delta)$ as suggested by \citet{chen2008extended}. The optimal $\lambda_{\text{opt}}$ is to minimize the EBIC
\begin{equation}
\label{opt: tuning}
\lambda_{\text{opt}} = \argmin_{\lambda} \text{EBIC}(\hat{\bbeta}^{n}_{\lambda}) .
\end{equation}
over  $\lambda_{1}, \lambda_{2}, \ldots$, with $\nu(\hat{\bbeta}^{n}_{\lambda_{i}}) = i$. 
Tunning parameter $\gamma $ in \eqref{eq:opt} is  chosen as $0.01 \lambda_{\text{opt}}$.  

\section{Applications to Gut Microbiome Studies}
\label{sec: IBD}
 The proposed method was applied to a study aiming at exploring the association between the pediatric inflammatory bowel disease  and gut microbiome conducted at the University of Pennsylvania  \citep{lewis2015inflammation}. 
This study collected the fecal samples of  85 IBD cases and 26 normal controls  and conducted a metagenomic sequencing for each sample, resulting a total of 97 bacterial species identified.  Among these bacterial species,  77 have non-zero values in at least 20 percent of the samples  were used in our analysis. The zero values in the relative abundance matrix were replaced with  0.5 times the minimum abundance observed, which is commonly used in microbiome data analyses  \citep{kurtz2015sparse, CaoLinLi}. Since the relative abundances of major species are relatively  large, replacing those zeros with a small value would not influence our results. The composition of species is then computed after replacing the zeros and used to fit the regression model.  

\subsection{Identifying bacterial species associated with IBD}
The proposed method was first applied to the logistic regression analysis between IBD and log-transformed compositions of the 77 species as covariates. To be specific, let $y$ be the binary indicator of IBD and   $\log (X_{k})$ is the logarithm of the relative abundance of the $k$-th species. We consider the following model 
\[
\text{logit}(Pr(y=1)) = \bbeta_{0} + \sum_{k=1}^{77} \bbeta_{k} \log (X_{k}),  \quad \text{where} \ \sum_{k=1}^{77} \bbeta_{k} =0.
\] 
Our goal is to  identify the bacteria species that are associated with IBD and to evaluate how well one can predict IBD based on the gut microbiome composition. 

\begin{figure}[h]
	\centering
	\begin{tabular}{cc}
	\includegraphics[height=0.25\textheight,width=0.49\textwidth]{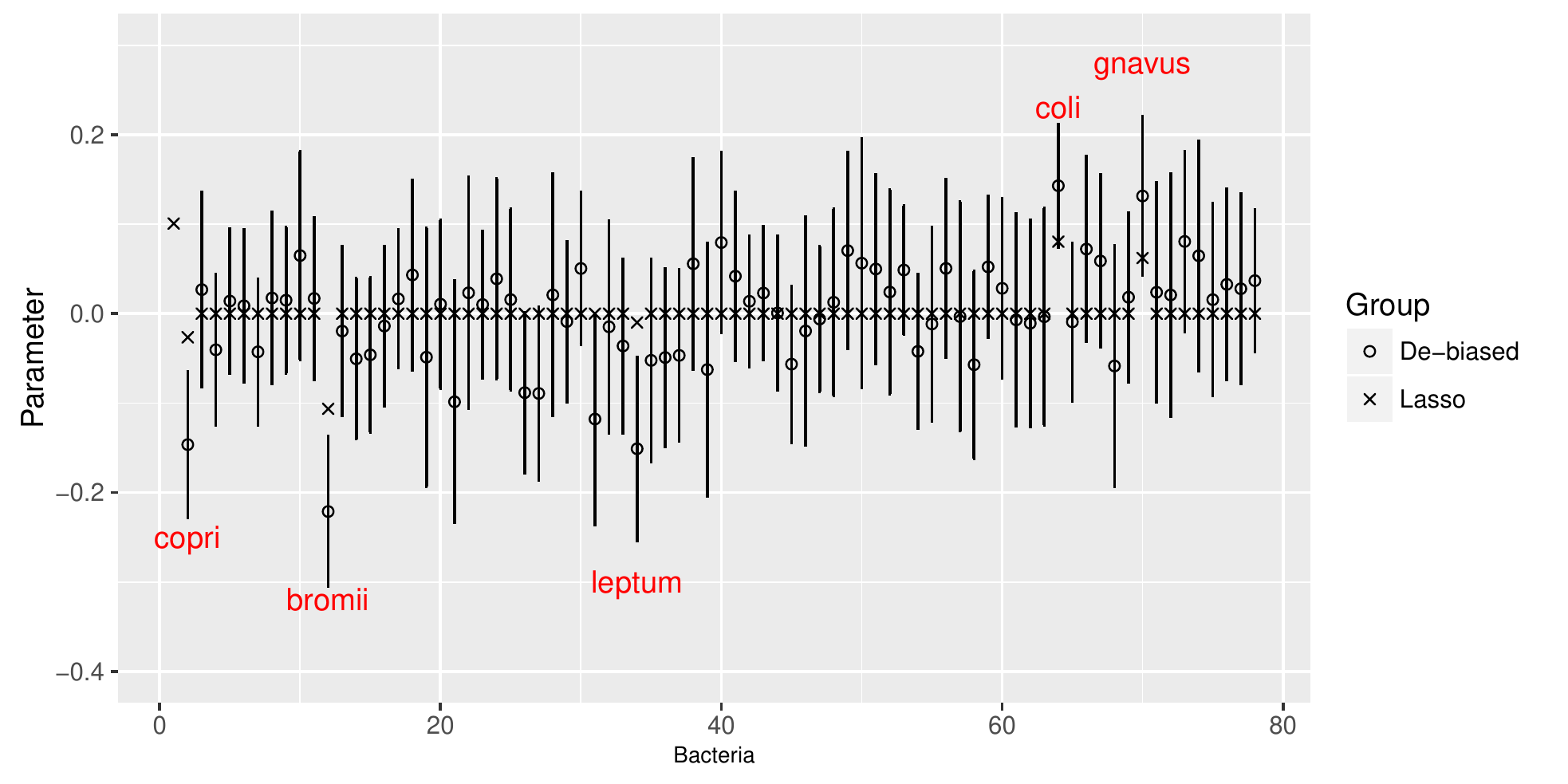} &
	\includegraphics[height=0.25\textheight,width=0.49\textwidth]{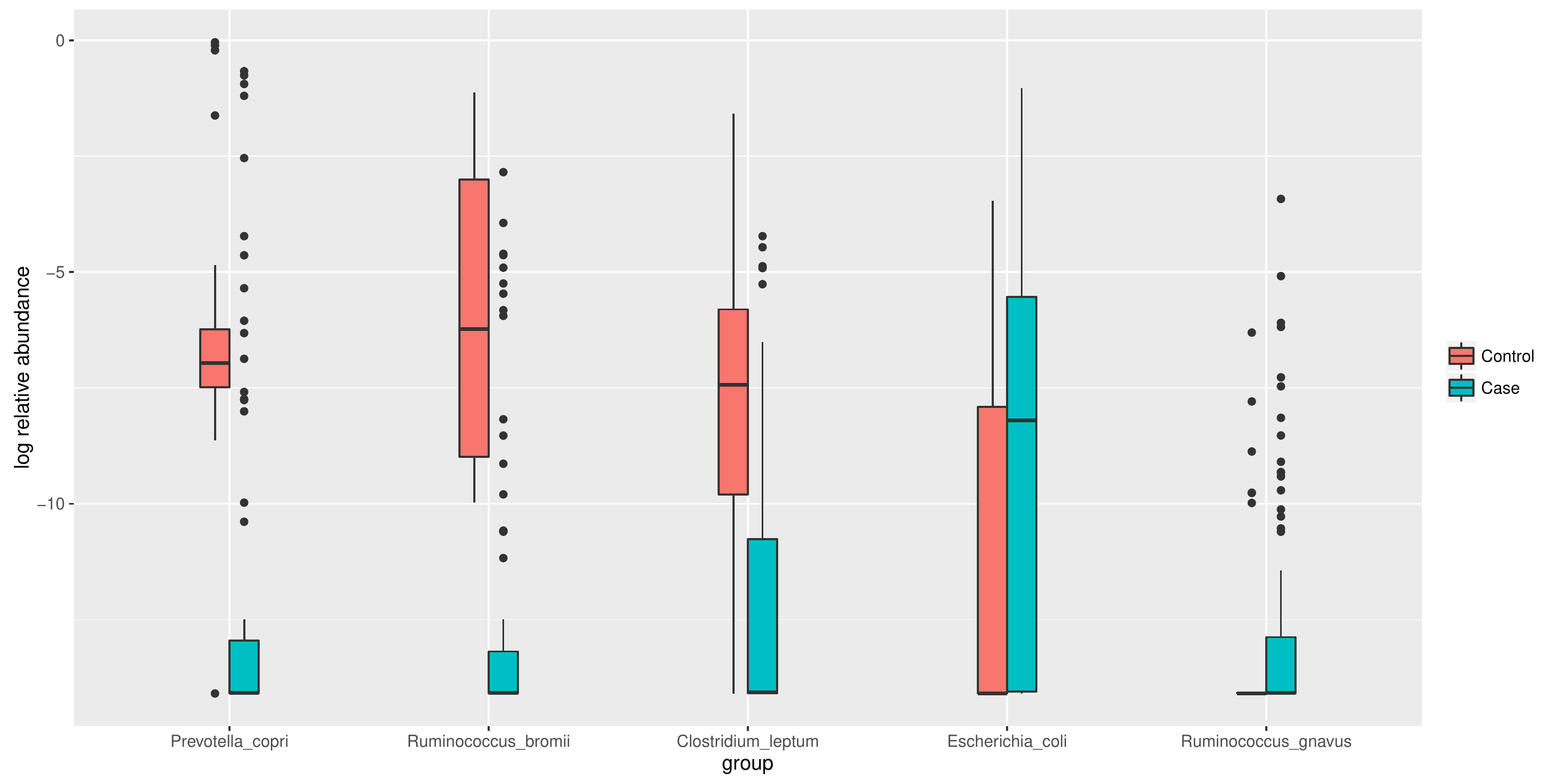}\\
	(a) & (b)\\
	\includegraphics[height=0.25\textheight,width=0.49\textwidth]{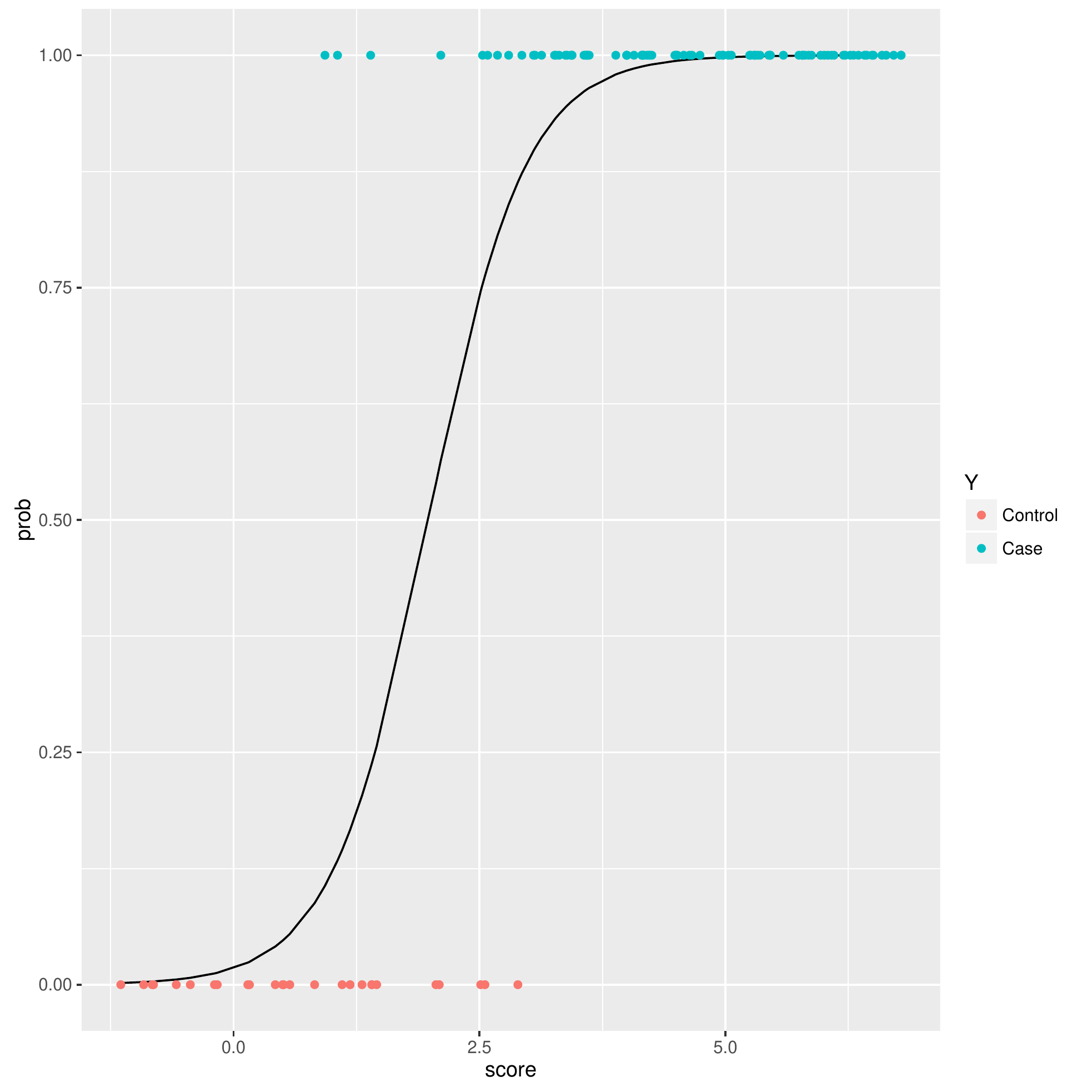} &
\includegraphics[height=0.25\textheight,width=0.49\textwidth]{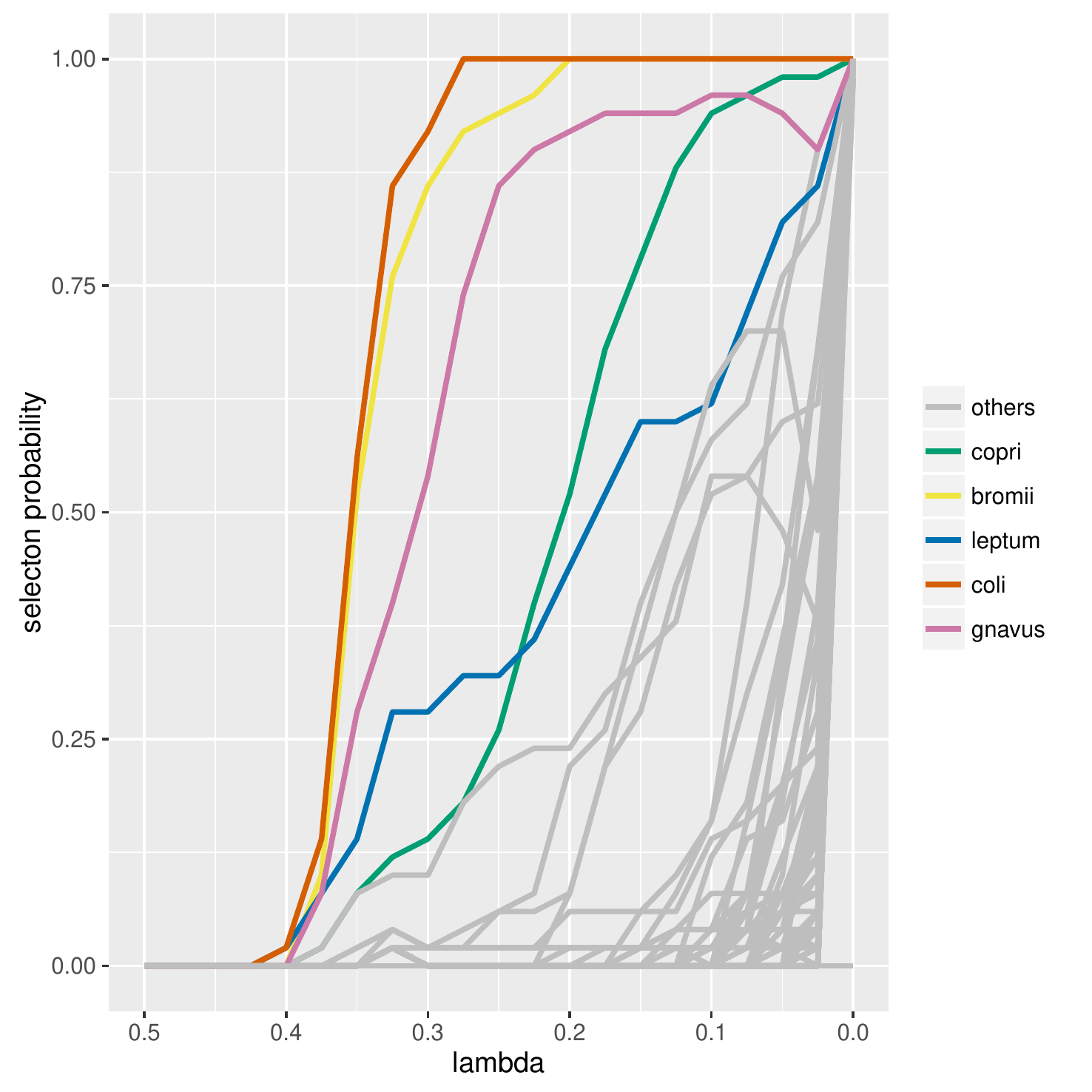}\\
(c) & (d)\\
\end{tabular}
\vspace{0.3in}
	\caption{Analysis of the IBD microbiome data. (a) Lasso estimates, de-biased estimates and $95 \%$ confidence intervals of the regression coefficients. Species selected based on  the CIs are annotated. (b)  Boxplots of log-relative abundances of the five identified species. The red and blue boxplots correspond to controls and cases samples respectively.
	(c) Fitted probability plot.  (d) Selection stability plot. 
}
	\label{IBD_onec}
\end{figure}  

Figure \ref{IBD_onec} (a) shows the Lasso estimates, de-biased estimates and $95 \%$ confidence intervals of the regression coefficients in the model. Five bacteria were selected using our methods with the 95\% CI not including zero,  including \emph{Prevotella\_copri}, \emph{Ruminococcus\_bromii}, \emph{Clostridium\_leptum}, \emph{Escherichia\_coli} and \emph{Ruminococcus\_gnavus}. The estimated coefficients and the corresponding 95\% CIs are summarized in Table \ref{Estimation: IBD}.   Among them, \emph{Prevotella\_copri}, \emph{Ruminococcus\_bromii}, \emph{Clostridium\_leptum} are negatively associated with the risk of IBD, indicating possible beneficial effects on  IBD.  On the other hand,  \emph{Escherichia\_coli} and \emph{Ruminococcus\_gnavus} are positively associated with IBD. 
  Figure \ref{IBD_onec} (b) plots the log-relative abundances of the five identified species in IBD children and in controls, indicating the the identified bacterial species indeed showed differential abundances between IBD cases and controls.

Our results  confirm the results from other studies.  \citet{kaakoush2012microbial} showed healthy people have high level of \emph{Prevotella\_copri} within their fecal microbial compared to Crohn's disease patients. \emph{Ruminococcus\_bromii} and \emph{Clostridium\_leptum} \citep{mondot2011highlighting, sokol2009low,kabeerdoss2013clostridium} were also  shown to be   negatively associated with  the risk of IBD. Furthermore, \citet{rhodes2007role} pointed  out the association of an increase of \emph{Escherichia\_coli} and IBD. \citet{matsuoka2015gut} also indicated the abundance of \emph{Ruminococcus\_gnavus} is higher in IBD patients.

\begin{table}[ht!]
	\caption{Selected bacteria and their corresponding phylum, estimated coefficients(standard errors in the parenthesis) and $95 \%$ confidence intervals.}	\label{Estimation: IBD}%
	\begin{center}
	\begin{tabular}{llr@{.}lc}
		\hline
		Bacteria name & Phylum &\multicolumn{2}{c}{$\bbeta$(se)}  & CI\\
		\hline 
		\emph{Prevotella\_copri} & Bacteroidetes  & $-0$&$15(0.042)$ & $(-0.23  ,-0.064)$\\
		\emph{Ruminococcus\_bromii} & Firmicutes & $-0$&$22(0.043)$  & $(-0.31 ,-0.18)$  \\
		\emph{Clostridium\_leptum} & Firmicutes & $-0$&$15(0.052)$  & $(-0.25, -0.048)$ \\
		\emph{Escherichia\_coli}  & Proteobacteria & $0$&$14(0.035) $ & $(0.074  , 0.21)$ \\
		\emph{Ruminococcus\_gnavus} & Firmicutes & $0$&$13(0.045)$  & $(0.043, 0.22)$  \\
		\hline 
	\end{tabular}
\end{center}
\end{table}

\ignore{
\begin{table}[ht!]
	\caption{Selected bacteria and their corresponding phylum, estimated coefficients(standard errors in the parenthesis) and $95 \%$ confidence intervals.}	\label{Estimation: IBD}%
	\begin{center}
		\begin{tabular}{llcc}
			\hline
			Bacteria name & Phylum &{$\bbeta$(se)}  & CI\\
			\hline 
			\emph{Prevotella\_copri} & Bacteroidetes  & $-0.15(0.042)$ & $(-0.23  ,-0.064)$\\
			\emph{Ruminococcus\_bromii} & Firmicutes & $-0.22(0.043)$  & $(-0.31 ,-0.18)$  \\
			\emph{Clostridium\_leptum} & Firmicutes & $-0.15(0.052)$  & $(-0.25, -0.048)$ \\
			\emph{Escherichia\_coli}  & Proteobacteria & $0.14(0.035) $ & $(0.074  , 0.21)$ \\
			\emph{Ruminococcus\_gnavus} & Firmicutes & $0.13(0.045)$  & $(0.043, 0.22)$  \\
			\hline 
		\end{tabular}
	\end{center}
\end{table}
}

\subsection{Stability, model fit and prediction evaluation}
To assess how stable the results are, we performed stability selection analysis  \citep{meinshausen2010stability} by sample splitting.  Among the 50 replications, each time we randomly sampled two third of the data including 56 cases and 16 controls and fit the model under different tuning parameters. Figure \ref{IBD_onec} (d) shows the selection probability for each of the bacteria versus  values of the tuning parameter. We see that the selected species in the previous section have the highest stability selection probabilities, indicating the 5 species selected are very stable.  Figure \ref{IBD_onec} (c)  shows the  fitted probability curve that  is constructed based on the five identified species,  indicating that our model fits the data well.

We then evaluate the performance of  prediction based on the IBD data. The data was randomly separated into a  training set  of 56 cases and 16 controls that  is used to estimate the parameters and  a testing set of 28 cases and 8 controls that is used to evaluate the prediction performance.  We  used the estimated parameters to predict the IBD status in the testing set and evaluated the performance based on area under the ROC curve (AUCs).  The procedure was repeated  50 times.    The	average AUC (se) are 0.92(0.049) , 0.93(0.043)  and  0.93 (0.051) based on Lasso, debiased  Lasso  and de-biased Lasso using only the selected bacterial species, indicating that the model can predict IBD very well.

\section{Simulation Studies}
\label{sec: simulation}
We evaluate the performance of of the proposed methods  through a set of simulation studies. In order to simulate covariate $Z$ and outcome $Y$, we simulate the true bacterial abundances  $W$, where  each row of $W$ is generated from  a log-normal distribution $\ln N(\mu, \bSigma)$, where $\bSigma_{ij}  = \zeta^{|i-j|}$ with $\zeta =0.2$ is the covariance matrix to reflect the correlation between different taxa. Mean parameters are set as  $\mu_{j} = \frac{p}{2}$ for $j= 1, \ldots,5$ and $\mu_{j} =1$ for $j=6, \ldots p$. The log-compositional covariate matrix $\bZ$ is obtained by normalizing the true abundances 
\[
\bZ_{ij} = \log \left(\frac{W_{ij}}{\sum_{k=1}^{p} W_{ik}}\right),
\] 
for $i= 1,2, \ldots, n$ and $j =1 ,2, \ldots, p$.  The true parameter $\bbeta$ is
\[
\bbeta = (0.45, -0.4, 0.45, 0, -0.5, 0, 0, 0, 0, 0, -0.6, 0, 0.3, 0, 0, 0.3, 0, \ldots 0)
\]
and $\bbeta_{0}=-1$. Based on these covariates, we simulate the binary outcome $Y$ based on the logistic probability $p_{i} = \text{expit}( \bZ_i^\top\bbeta+ \bbeta_{0})$ and obtained  the number of cases and controls at a 2:3 ratio.  Different dimensions and sample sizes are considered and simulations are repeated 100 times for each setting. The true regression coefficients $\bbeta$  are assumed to satisfy the following linear  constraints:
\begin{equation*}
\begin{aligned}
\sum_{i=1}^{10} \bbeta_{i}=0, \sum_{i=11}^{16} \bbeta_{i}=0, \sum_{i=17}^{20} \bbeta_{i}=0, \sum_{i=21}^{23} \bbeta_{i}=0,\\
\sum_{i=24}^{30} \bbeta_{i}=0, \sum_{i=31}^{32} \bbeta_{i}=0, \sum_{i=33}^{40} \bbeta_{i}=0, \sum_{i=41}^{p} \bbeta_{i}=0.
\end{aligned}
\end{equation*}

\subsection{Simulation results}
We evaluate the performance of the simulation by comparing the coverage probability, length of the confidence interval and the true positive and false positive of selecting variables based on the confidence interval. We compare the results of fitting the models with no constraint, one constraint, true constraint and 
misspecified constraints specified  below,
\begin{equation*}
\begin{aligned}
\sum_{i=1}^{4} \bbeta_{i}=0, \sum_{i=5}^{12} \bbeta_{i}=0, \sum_{i=13}^{23} \bbeta_{i}=0, \sum_{i=24}^{30} \bbeta_{i}=0,
\sum_{i=31}^{p} \bbeta_{i}=0.
\end{aligned}
\end{equation*}

\begin{figure}[h]
	\centering
	\begin{tabular}{cc}
	\includegraphics[height=0.25\textheight,width=0.49\textwidth]{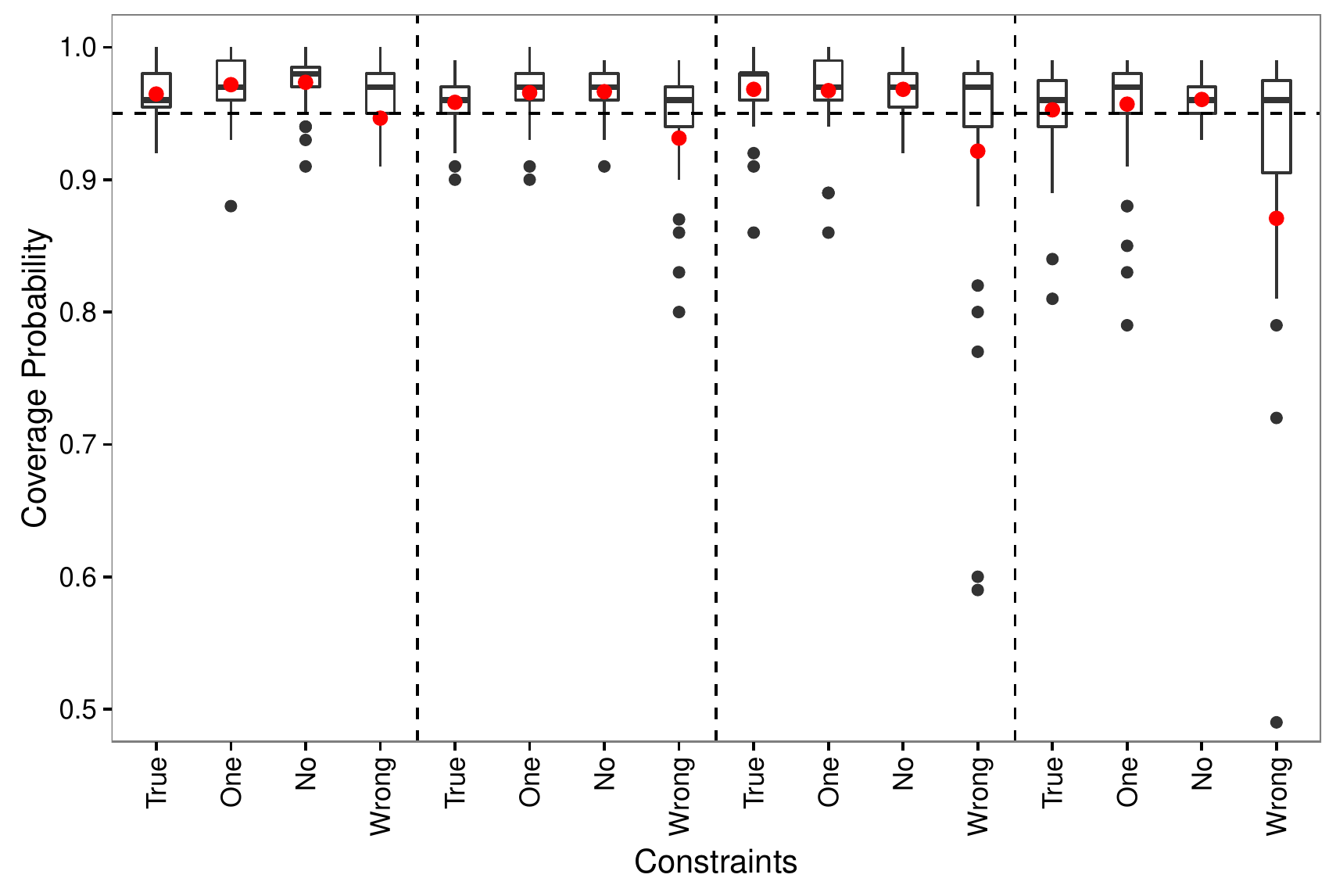} &
	\includegraphics[height=0.25\textheight,width=0.49\textwidth]{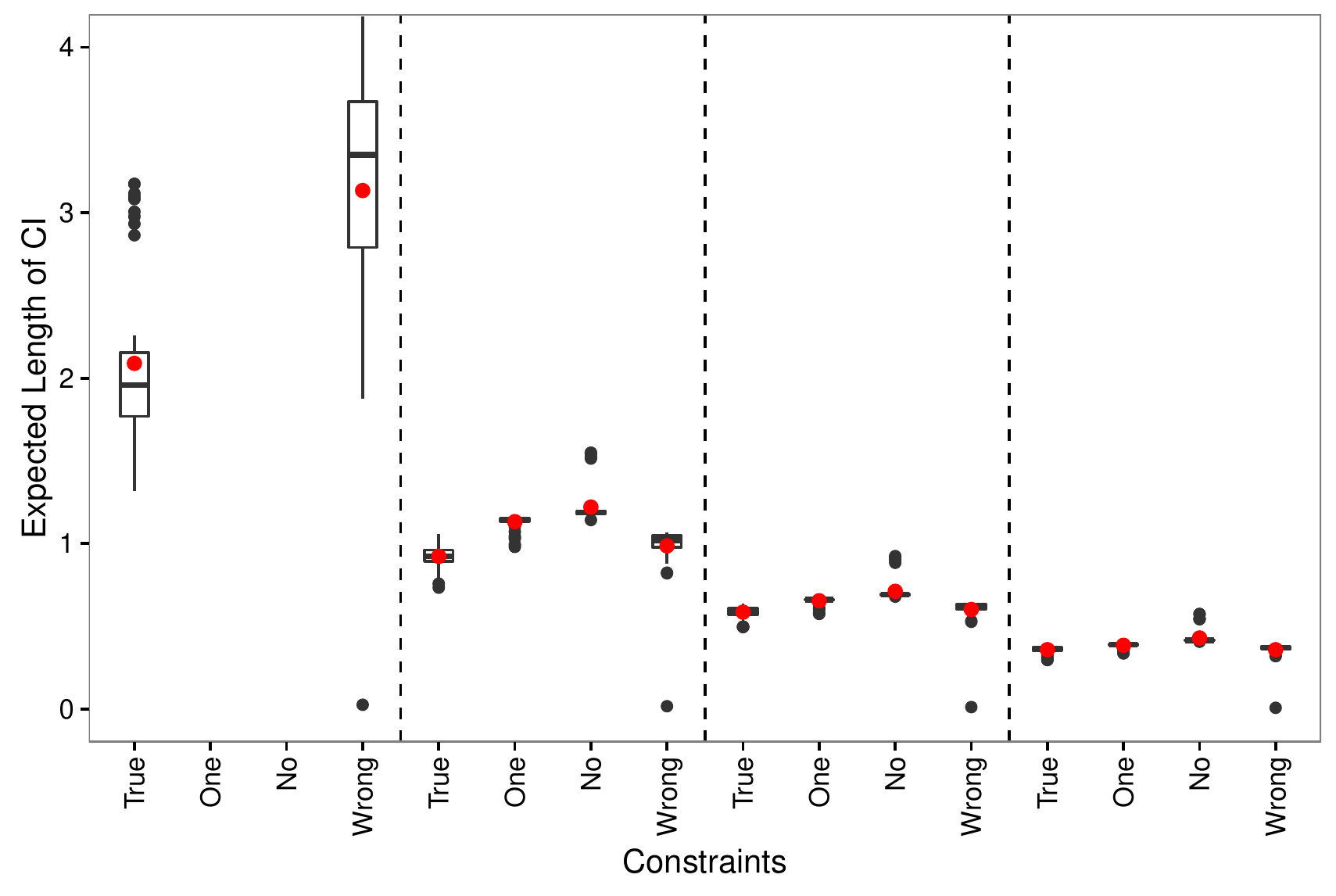} \\
	(a) & (b)\\
	\\
	\includegraphics[height=0.25\textheight,width=0.49\textwidth]{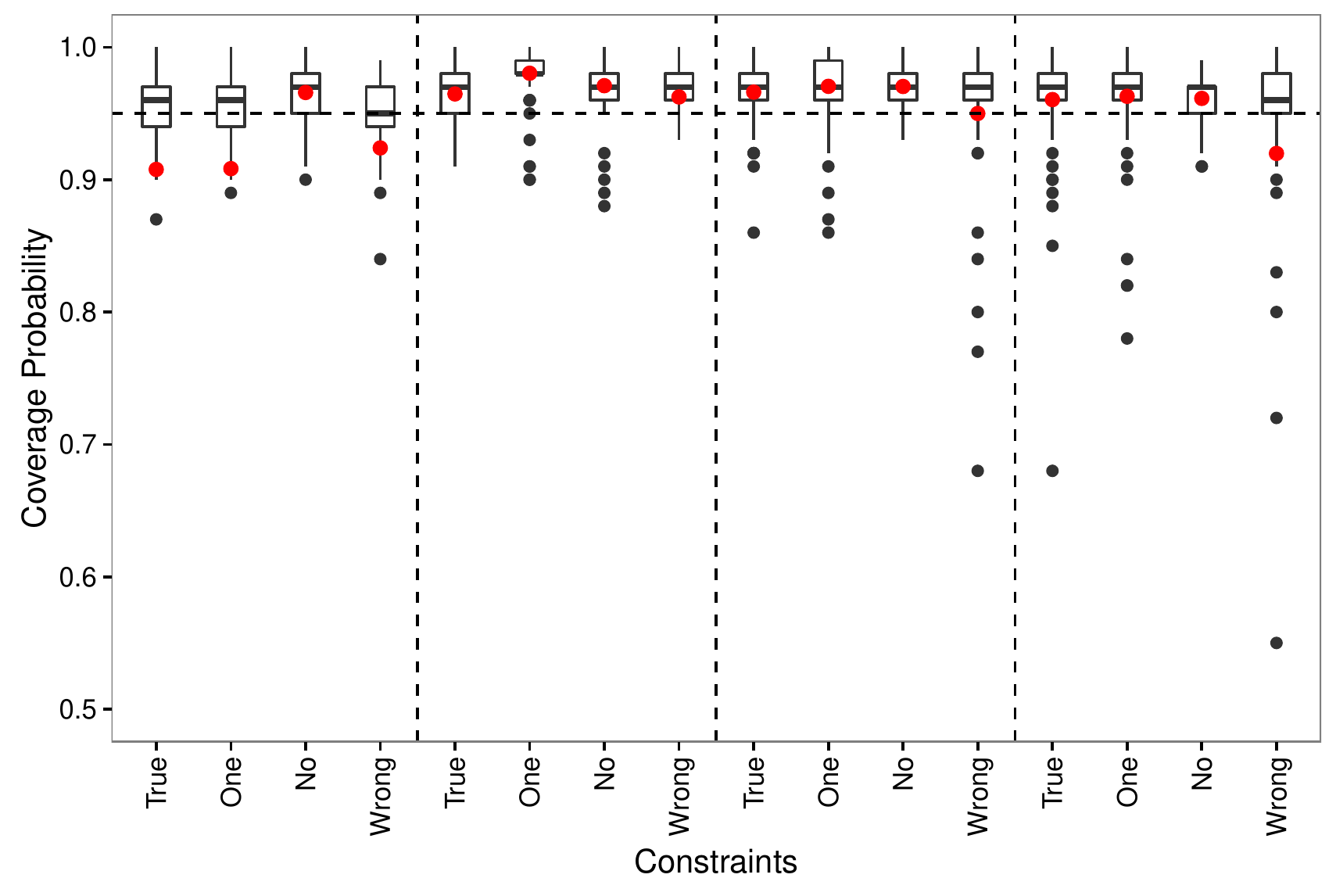} &
	\includegraphics[height=0.25\textheight,width=0.49\textwidth]{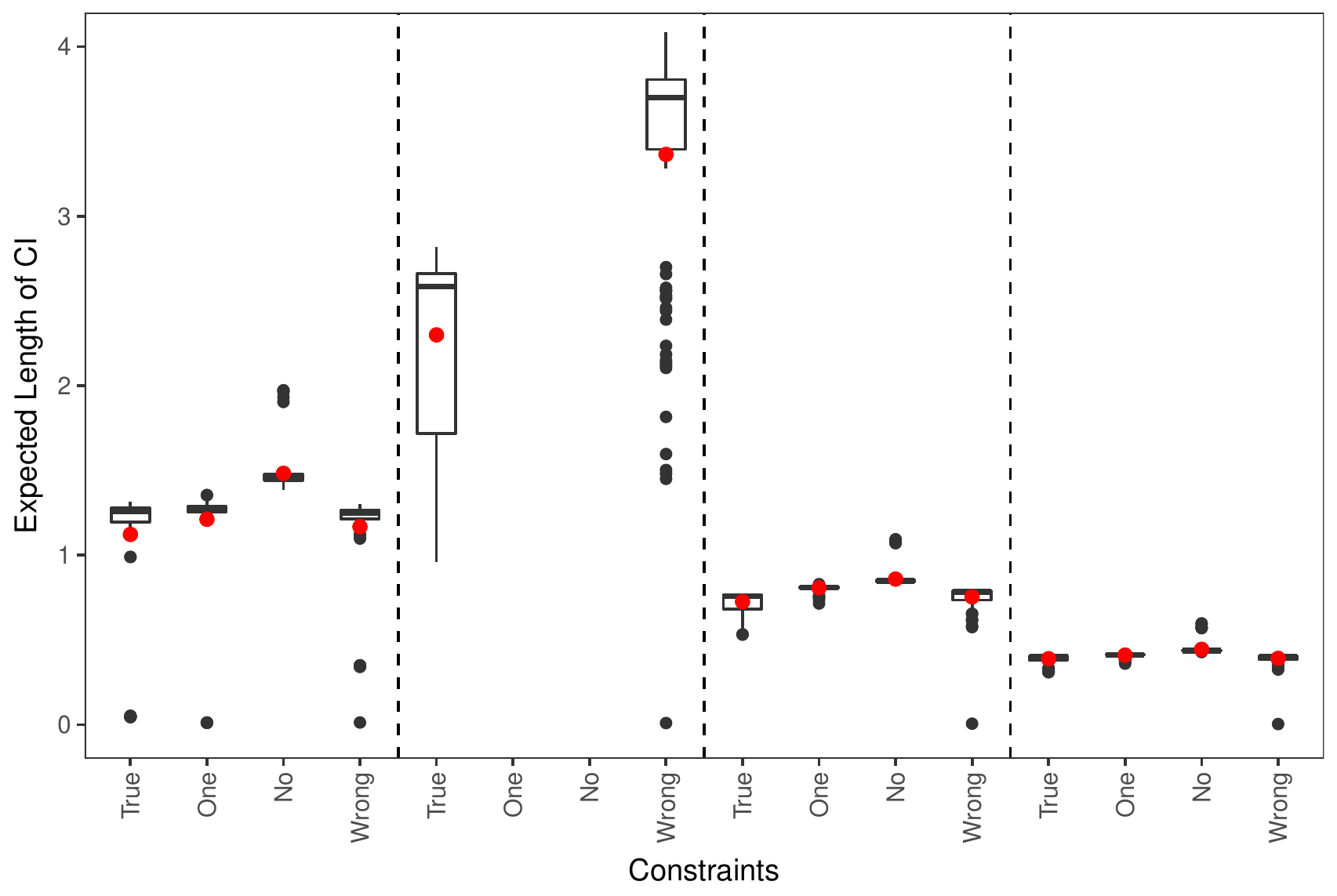} \\
	(c) & (d)\\
	\\
	\\
	\end{tabular}
	\caption{Coverage probabilities and length of confidence intervals based on 100 simulations for $p=50$ ((a) and (b)) and $p=100$ ((c) and (d)) and  $n = 50,100,200,500$ (separated by vertical dashed lines).}
	\label{fig:sim_generated_cov}
\end{figure}

Figure \ref{fig:sim_generated_cov} shows  that the coverage probabilities are closer  to $95 \%$  and the length of CIs decrease as sample size becomes larger. In addition, the coverage probabilities under true constraints are closer to the correct coverage probability ($95\%$) especially when $n$ is relatively larger($n=200, 500$). As for length of CIs,  the CIs using the true constraints have the shortest CIs while the length of the CIs for single constraint and no constraints are relatively wider. We did not compare the length of CI for using misspecified constraints because the coverage probability in this case is really poor.  The figure also shows that the coverage probabilities are sensitive to the constraints when sample size becomes larger and the length is sensitive to the constraints for small sample size. This is expected as when the sample size is small, we are more likely to obtain wider CI, and using the correct constraints, which provide more information, would provide shorter CI. While for the coverage probability, since our algorithm provides an asymptotic CI,  the sample size has bigger effects than the constraints.  The coverage probability becomes really poor when   the constraints are misspecified  when $n=500$.  

Table \ref{table:T&F rate} shows the true positive and false positive rates of selecting the significant variables using the $95\%$ confidence interval under multiple, one, no and misspecified constraints for various dimensions $p$ and sample sizes $n$.  The false positive rates are correctly controlled under $5\%$ for all models, even when the constraints are misspecified. However, models with correctly specified  linear constraints have higher true positive rates. When the sample size is 500,  true positive rate is greater  than $90\%$,  which is the highest among all models  considered.

\begin{table}[htbp]
	\centering
	\caption{True /False positive rates of the significant variables selected by the $95\%$ confidence interval using multiple, one, no and misspecified constraints. $p=50, 100$ and  $n=50,100,200,500$ are considered. }	\label{table:T&F rate}%
\begin{tabular}{ccccccccccccc}
		\hline
		$n$    & & TP    & FP    && TP    & FP    && TP    & FP   & & TP    & FP \\
		\cline{1-1}  \cline{3-4} \cline{6-7} \cline{9-10} \cline{12-13}
		&&\multicolumn{2}{c}{Multi}& & \multicolumn{2}{c}{One} &&\multicolumn{2}{c}{No}&&\multicolumn{2}{c}{Wrong} \\
		\cline{1-1}  \cline{1-1}  \cline{3-4} \cline{6-7} \cline{9-10} \cline{12-13}
		\multicolumn{13}{c}{$p=50$}\\
		50   & & 0.069  & 0.034  && 0.026  & 0.025  && 0.029  & 0.026 & & 0.054  & 0.036  \\
		100   && 0.260  & 0.038  && 0.206  & 0.031 & & 0.141  & 0.034  && 0.299  & 0.038  \\
		200  & & 0.569  & 0.026  && 0.549  & 0.025  && 0.411  & 0.030  && 0.546  & 0.037  \\
		500   && 0.914  & 0.038  && 0.897  & 0.030  && 0.840  & 0.038  && 0.814  & 0.058  \\
			\multicolumn{13}{c}{$p=100$}\\
	50    && 0.220  & 0.045  && 0.071  & 0.044  && 0.109  & 0.034  && 0.134  & 0.046  \\
100   &&0.103  & 0.035  && 0.023  & 0.016  && 0.107  & 0.026  && 0.154  & 0.027  \\
200   && 0.431  & 0.030  && 0.389  & 0.025  && 0.283  & 0.029  && 0.481  & 0.032  \\
500   && 0.907  & 0.032  && 0.873  & 0.029  && 0.801  & 0.037  && 0.804  & 0.042  \\			
		\hline
\end{tabular}
\end{table}

\ignore{
\begin{table}[htbp]
	\centering
	\caption{True /False positive rates of the significant variables selected by the $95\%$ confidence interval using multiple, one, no and misspecified constraints. Different $n$ is considered and $p=100$.}	\label{table: T&F rate_100}%
	\begin{tabular}{ccccccccccccc}
		\hline
		$n$    & & TP    & FP    && TP    & FP    && TP    & FP   & & TP    & FP \\
		\cline{1-1}  \cline{3-4} \cline{6-7} \cline{9-10} \cline{12-13}
		&&\multicolumn{2}{c}{Multi}& & \multicolumn{2}{c}{One} &&\multicolumn{2}{c}{No}&&\multicolumn{2}{c}{Wrong} \\
		\cline{1-1}  \cline{1-1}  \cline{3-4} \cline{6-7} \cline{9-10} \cline{12-13}
		50    && 0.220  & 0.045  && 0.071  & 0.044  && 0.109  & 0.034  && 0.134  & 0.046  \\
		100   &&0.103  & 0.035  && 0.023  & 0.016  && 0.107  & 0.026  && 0.154  & 0.027  \\
		200   && 0.431  & 0.030  && 0.389  & 0.025  && 0.283  & 0.029  && 0.481  & 0.032  \\
		500   && 0.907  & 0.032  && 0.873  & 0.029  && 0.801  & 0.037  && 0.804  & 0.042  \\
		\hline
		\end{tabular}
\end{table}
}

\section{Discussion}
\label{sec: discussion}
We have considered  estimation and inference for the generalized linear models with high dimensional compositional covariates. In order to accounting for the nature of compositional data, a group of linear constraints are imposed on the regression coefficients to ensure subcompositional coherence.
With these constraints, the standard GLM Lasso algorithm based on Taylor expansion and coordinate descent algorithm does not work due to the non-separable nature of the penalty function. Instead, a generalized accelerated proximal gradient algorithm was developed to estimate the regression coefficients.   To make statistical inference,  a de-biased procedure is proposed to construct valid confidence intervals of the regression coefficients, which could be used for hypothesis testing as well as identifying species that are associated with the outcome. Application of the method to an analysis of IBD microbiome data has identified five bacterial species that are associated with pediatric IBD with a high stability. The identified model has also shown a great prediction performance based on cross-validation. 

The approach we took in deriving  the confidence intervals follows that of \citet{javanmard2014confidence} by first obtaining an debiased estimates of the regression coefficients.   Alternatively, one can consider the approach based on post-selection  inference for $\ell_1$-penalized likelihood models \citep{Taylor}. However, one needs to modify the methods for  \cite{Taylor} to take into account the linear constraints of the regression coefficients. It would be interesting to compare the performance of this alternative approach. 

\section*{Appendix}
We provide proofs for the main theorems in the paper.

\begin{lemma}
	If Conditions C1 and C2 hold, then for any matrix $A$, 
	\[
	|(\biden - P_C)A|_{\infty} \leq k_0 |A|_{\infty}.
	\]
\end{lemma}
The proof for this lemma is in the appendix of \citet{shi2016}.


\subsection*{Proof of Theorem \ref{thm: consistency}}
\begin{proof}
	By the definition of $\hat{\bbeta}^{n}$ and (\ref{eq:L1}), we have:
	\begin{equation} \label{eq: consis}
	-\dfrac{1}{n}[Y^\top \btZ\hat{\bbeta}^n - \sum_{i=1}^nA(\widetilde{Z}_i^\top\hat{\bbeta}^n)]+\lambda||\hat{\bbeta}^n||_1 \leq  -\dfrac{1}{n}[Y^\top \btZ\bbeta - \sum_{i=1}^nA(\widetilde{Z}_i^\top\bbeta)]+\lambda||\bbeta||_1.
	\end{equation}
	Denote $h = \hat{\bbeta}^n - \bbeta$,  and $S_{h}$ be the set of index of the $s$ largest absolute values of $h$. Then rearrange (\ref{eq: consis}), we get:
	\begin{equation}
	\label{ineq: lambda}
	\lambda(\|\bbeta\|_{1} - \|\hat{\bbeta}^{n}\|_{1}) \geq -\dfrac{1}{n}[Y^\top \btZ h - \sum_{i=1}^n(A(\widetilde{Z}_i^\top\hat{\bbeta}^n)-A(\widetilde{Z}_i^\top\bbeta))].
	\end{equation}
	Notice that,
	\begin{align}
	\label{ineq: beta1}
	\|\bbeta\|_{1} - \|\hat{\bbeta}^{n}\|_{1}  =& \|\bbeta_{supp(\bbeta)}\|_{1} - \|\hat{\bbeta}_{supp(\bbeta)}^{n}\|_{1} - \|\hat{\bbeta}_{supp(\bbeta)^{c}}^{n}\| _{1}, \nonumber \\
	\leq & \|\bbeta_{supp(\bbeta)}- \hat{\bbeta}_{supp(\bbeta)}^{n}\|_{1}- \|h_{supp(\bbeta)^{c}}\|_{1},  \nonumber \\
	\leq &\|h_{S_{h}}\|_{1}  -\|h_{S^{c}_{h}}\|_{1}.
	\end{align}
Furthermore, for each $i$ applied the mean value theorem to $A$ defined in \ref{eq:exp_fam}, there exists $\widetilde{\beta_{i}}^0$ such that $A(\widetilde{Z}_i^\top\hat{\beta}^n)-A(\widetilde{Z}_i^\top\beta) =\mu(\widetilde{\beta}, \widetilde{Z}_i)\widetilde{Z}_i^\top h + \dfrac{1}{2}v(\widetilde{\beta_{i}}^0,\widetilde{Z}_i)\left(\widetilde{Z}_i^\top h\right)^2$. Then we have:
\begin{align}
& -\dfrac{1}{n}[Y^\top \btZ h - \sum_{i=1}^n(A(\widetilde{Z}_i^\top\hat{\beta}^n)-A(\widetilde{Z}_i^\top\beta))]\\
& \geq   -\dfrac{1}{n}[Y^\top \btZ h - \mu(\beta, \btZ)^\top \btZ h)], \nonumber\\
& \geq  -\dfrac{1}{n}( Y -\mu(\beta, \btZ))^\top \btZ h, \nonumber\\
& \geq -\dfrac{1}{n}\| Y -\mu(\beta, \btZ)^\top \btZ \|_{\infty} \cdot \|h\|_{1}  =   -\dfrac{1}{n}\|( Y -\mu(\beta, \btZ))^\top \btZ \|_{\infty} \cdot (\|h_{S_{h}}\|_{1}  +\|h_{S^{c}_{h}}\|_{1}). \nonumber
\end{align} 
When the event $\| (Y -\mu(\beta, \btZ))^\top \btZ \|_{\infty} \leq \dfrac{n\lambda}{\tau}$ holds, we have:
	\begin{equation}
	\label{ineq: beta2}
	\lambda(\|\bbeta\|_{1} - \|\hat{\bbeta}^{n}\|_{1}) \geq  -\dfrac{1}{n} \cdot \frac{n\lambda}{\tau} \cdot (\|h_{S_{h}}\|_{1}  +\|h_{S^{c}_{h}}\|_{1}).
	\end{equation}
	So by \eqref{ineq: lambda}, \eqref{ineq: beta1} and \eqref{ineq: beta2} we have:
	\begin{align*}
	\lambda(\|h_{S_{h}}\|_{1}  -\|h_{S^{c}_{h}}\|_{1}) \geq \lambda (\|\bbeta\|_{1} - \|\hat{\bbeta}^{n}\|_{1}) \geq    - \frac{\lambda}{\tau} \cdot (\|h_{S_{h}}\|_{1}  +\|h_{S^{c}_{h}}\|_{1}).
	\end{align*}
	That is, 
	\begin{equation}
	\label{ineq:hc}
	\|h_{S^{c}_{h}}\|_{1} \leq \dfrac{\tau+1}{\tau-1}\|h_{S_{h}}\|_{1}.
	\end{equation}
	Then by the KKT condition of optimization problem \eqref{eq:L1}, we have:
	\begin{equation}
	\| \btZ^\top (Y -\mu(\hat{\bbeta}^{n}, \btZ)) + C\bm{\eta}\|_{\infty} \leq n\lambda,
	\end{equation}
	for some $\bm{\eta} \in \R^{r}$. Then by Lemma 1, 
	\begin{align}
	\| (\biden- P_C)\left(\btZ^\top (Y -\mu(\hat{\bbeta}^{n}, \btZ)) + C\bm{\mu}\right)\|_{\infty} & \leq k_{0} \| \btZ^\top (Y -\mu(\hat{\bbeta}^{n}, \btZ)) + C\bm{\mu}\|_{\infty} \leq k_0 n \lambda. 
	\end{align}
	Then as
	\begin{align*}
	(\biden- P_C)(\btZ^\top (Y -\mu(\hat{\bbeta}^{n}, \btZ)) + C\bm{\mu}) & = (\biden- P_C)\btZ^\top (Y -\mu(\hat{\bbeta}^{n}, \btZ)) + (\biden- P_C)C\bm{\mu}, \\
	& = \btZ^\top (Y -\mu(\hat{\bbeta}^{n}, \btZ)).
	\end{align*}
	with the the assumption that $\| (Y -\mu(\bbeta, \btZ))^\top \btZ \|_{\infty} \leq \dfrac{n\lambda}{\tau}$, we have:
	\begin{align}
	\|\btZ^\top(\mu(\hat{\bbeta}^{n}, \btZ) - \mu(\bbeta, \btZ))\|& \leq  \|\btZ^\top (Y -\mu(\hat{\bbeta}^{n}, \btZ)) \|_{\infty}+\|\btZ^\top (Y -\mu(\bbeta, \btZ)) \|_{\infty} \leq k_0 n \lambda + \dfrac{n\lambda}{\tau}. \nonumber
	\end{align}
	As $\|\btZ^\top(\mu(\hat{\bbeta}^{n}, \btZ) - \mu(\bbeta, \btZ))\| = \|\btZ^\top \mathbf{V}(\bbeta^{0},\btZ) \btZ h \|_{\infty}$,  we get
	\[\|\btZ^\top \mathbf{V}(\bbeta^{0},\btZ) \btZ h \|_{\infty} \leq k_0 n \lambda + \dfrac{n\lambda}{\tau} . \]
	Since $\mathbf{V}(\bbeta^{0},\btZ)$ is a diagonal matrix with all its nonzero elements greater than zero, define $\btZ_{v} = \mathbf{V}^{\frac{1}{2}}(\bbeta^{0},\btZ)\btZ $,  where $\mathbf{V}^{\frac{1}{2}}(\bbeta^{0},\btZ) = \diag\{(v(\bbeta^{0}, Z_1))^{\frac{1}{2}},\dots,(v(\bbeta^{0}, Z_n))^{\frac{1}{2}}\}$. So $\btZ_{v}^\top\btZ_{v} = \btZ^\top \mathbf{V}(\bbeta^{0},\btZ) \btZ$.  
	Using Lemma 5.1 in \citet{cai2013compressed}, we have: 
	\begin{align*}
	|\langle \btZ_{v} h_{S_{h}}, \btZ_{v} h_{S^{c}_{h}} \rangle|  &\leq \theta_{s,s}(\btZ_{v}) \|h_{S_{h}}\|_2 \cdot \max(\|h_{S^{c}_{h}}\|_{\infty}, \|h_{S^{c}_{h}}\|_{1}/s)\sqrt{s}, \\
	& \leq \sqrt{s}\theta_{s,s}(\btZ_{v})\|h_{S_{h}}\|_2  \cdot \frac{\tau +1}{\tau -1} \|h_{S_{h}}\|_{1} /s, \\
	& \leq \dfrac{\tau +1}{\tau -1}\theta_{s,s}(\btZ_{v})\|h_{S_{h}}\|^{2}_2.   
	\end{align*}
	Then, 
	\begin{align}
	\label{ineq:hSh}
	(k_0 n \lambda + \dfrac{n\lambda}{\tau})\|h_{S_{h}}\|_1 & \geq \|\btZ^\top \mathbf{V}(\bbeta^{0},\btZ) \btZ h \|_{\infty} \|h_{S_{h}}\|_1 \geq \langle \btZ_{v}^\top \btZ_{v}h, h_{S_{h}}\rangle,  \nonumber\\
	& = \langle \btZ_{v} h_{S_{h}},\btZ_{v} h_{S_{h}}\rangle +\langle \btZ_{v} h_{S_{h}},\btZ_{v} h_{S^{c}_{h}}\rangle, \nonumber\\
	& \geq \|\btZ_{v} h_{S_{h}}\|_{2}^{2} - \dfrac{\tau +1}{\tau -1}\theta_{s,s}(\btZ_{v})\|h_{S_{h}}\|^{2}_2, \nonumber\\
	& \geq \left( \delta_{2s}^{-}(\btZ_{v}) - \dfrac{\tau +1}{\tau -1}\theta_{s,s}(\btZ_{v})\right)\|h_{S_{h}}\|^{2}_2,  \nonumber\\ 
	& \geq \left(\dfrac{3\tau-1}{2(\tau-1)}\delta_{2s}^{-}(\btZ_{v}) -\dfrac{\tau+1}{2(\tau-1)}\delta_{2s}^{+}(\btZ_{v}) \right)\|h_{S_{h}}\|^{2}_1 /s.
	\end{align}
	So from \eqref{ineq:hSh} we have:
	\begin{align}
	\label{ineq:hSh2}
	\|h_{S_{h}}\|_1 & \leq \dfrac{s\left(k_0 n \lambda + \dfrac{n\lambda}{\tau}\right)}{\left(\dfrac{3\tau-1}{2(\tau-1)}\delta_{2s}^{-}(\btZ_{v}) -\dfrac{\tau+1}{2(\tau-1)}\delta_{2s}^{+}(\btZ_{v}) \right)}, \nonumber \\
	& \leq s\dfrac{k_0 n \lambda + \dfrac{n\lambda}{\tau}}{2n\tau\phi_{0}/(\tau -1)}. 
	\end{align} 
	So combine \eqref{ineq:hc} and \eqref{ineq:hSh2}, we have:
	\[
	\|\hat{\bbeta}^{n} - \bbeta\|_1 = \|h_{S_{h}}\|_1+\|h_{S^{c}_{h}}\|_1 \leq \dfrac{2\tau}{\tau -1}\|h_{S_{h}}\|_{1} \leq \dfrac{s\lambda(k_0+1/\tau)}{\phi_{0}} .
	\]
	Take $\lambda = \tau \tilde{c}\sqrt{(\log p)/n}$, so we have:
	\begin{align*}
	\P\left(\|\hat{\beta}^{n} - \beta\| _1  \leq \dfrac{s\lambda(k_0+1/\tau)}{\phi_{0}} \right) & \geq 1- \P\left( \| (Y -\mu(\beta, \btZ))^\top \btZ \|_{\infty} > \dfrac{n\lambda}{\tau}\right) \\
	& \geq 1-  \sum_{i=1}^{p} \P\left( | ((Y -\mu(\beta, \btZ))^\top \btZ )_{i}| > \dfrac{n\lambda}{\tau}\right) \\
	& \geq 1 -2 \sum_{i=1}^{p}\exp \left( - \dfrac{(\sqrt{n} \lambda / \tau}{2K^{2}}  \right)\geq 1 -2p^{1 - \tilde{c}^{2}/(2K^2)}
	\end{align*}
\end{proof}

\subsection*{Proof of Lemma \ref{thm: feasible set}}
\begin{proof}
We first provide a bound for $\bSigma$.
	Notice that:
	\begin{align*}
	\bm{\Omega}_{\bbeta}\bSigma -(\biden-P_C) &= \dfrac{1}{n}\sum_{k=1}^{n}\left( \bm{\Omega}_{\bbeta} v(\bbeta, \widetilde{Z}_{k})\widetilde{Z}_{k}\widetilde{Z}_{k}^\top  - (\biden-P_C)\right),  \\
	& = \dfrac{1}{n}\sum_{k=1}^{n}\left( \bm{\Omega}_{\bbeta}^{1/2}\bm{\Omega}_{\bbeta}^{1/2} v(\bbeta, \widetilde{Z}_{k})\widetilde{Z}_{k}\widetilde{Z}_{k}^\top\bm{\Omega}^{1/2}_{\bbeta} \bSigma_{\bbeta}^{1/2} - (\biden-P_C)\right).
	\end{align*}
	The last equality is true as $\bSigma_{\bbeta}^{1/2}\bm{\Omega}^{1/2}_{\bbeta}\widetilde{Z}_{k} = (\biden - P_C)\widetilde{Z}_{k} = \widetilde{Z}_{k}$ for $k =1,2 \ldots,n$. Then notice that $\E \bm{\Omega}_{\bbeta} v(\bbeta, \widetilde{Z}_{k})\widetilde{Z}_{k}\widetilde{Z}_{k}^\top = \E \bm{\Omega}_{\bbeta} \bSigma_{\bbeta} = \biden - P_C$, so define:
	\[
	v_{k}^{(ij)}  = \bm{\Omega}_{i,\cdot}^{1/2}\bm{\Omega}_{\bbeta}^{1/2} v(\bbeta, \widetilde{Z}_{k})\widetilde{Z}_{k}\widetilde{Z}_{k}^\top\bm{\Omega}^{1/2}_{\bbeta} (\bSigma_{\bbeta})_{\cdot,j}^{1/2} - (\biden-P_C)_{i,j},
	\]
	we know that  $\E v_{k}^{(ij)} =0$ for $k =1,2 \ldots,n$ and any $i,j$. Then by the proof of Lemma 6.2 in \citet{javanmard2014confidence}, we have:
	\begin{align*}
	\|v_{k}^{(ij)}\|_{\psi_{1}}  & \leq 2\| \bm{\Omega}_{i,\cdot}^{1/2}\bm{\Omega}_{\bbeta}^{1/2} v(\bbeta, \widetilde{Z}_{k})\widetilde{Z}_{k}\widetilde{Z}_{k}^\top\bm{\Omega}^{1/2}_{\bbeta} (\bSigma_{\bbeta})_{\cdot,j}^{1/2} \|_{\psi_{1}},   \\
	& \leq 2v(\bbeta, \widetilde{Z}_{k}) \| \bm{\Omega}_{i,\cdot}^{1/2}\bm{\Omega}_{\bbeta}^{1/2} \widetilde{Z}_{k}\|_{\psi_{2}} \|(\bSigma_{\bbeta})_{\cdot,j}\bm{\Omega}^{1/2}_{\bbeta}\widetilde{Z}_{k}  \|_{\psi_{2}}, \\
	& \leq 2 \|(\bSigma_{\bbeta})_{\cdot,j}\|_{2} \|\bm{\Omega}_{i,\cdot}^{1/2}\|_{2} \cdot \|\bm{\Omega}_{\bbeta}^{1/2} \widetilde{Z}_{k}\|_{\psi_{2}} \|\bm{\Omega}^{1/2}_{\bbeta}\widetilde{Z}_{k} \|_{\psi_{2}},  \\
	& \leq 2\sqrt{C_{\text{max}}/ C_{\text{min}}}\kappa^2 \equiv \kappa_{1}^{\prime}.
	\end{align*}
	Then by inequality for  centered sub-exponential random variables from \citet{buhlmann2011statistics}, we have:
	\[
	\P\left( \dfrac{1}{n} |\sum_{k=1}^{n}v_{k}^{(ij)}| \geq  \gamma \right) \leq \exp\left(  -\dfrac{n}{6} \min \left\{ \left(\dfrac{\gamma}{e\kappa^{'}}\right)^{2}, \left(\dfrac{\gamma}{e\kappa^{'}}\right)\right\}\right).
	\]
	Pick $\gamma = c \sqrt{(\log p) / n}$ with $ c \leq e \kappa_{1}^{\prime} \sqrt{n / (\log p)}$, we have:
	\begin{equation}
	\label{ineq:Bernstein}
	\P\left( \dfrac{1}{n} |\sum_{k=1}^{n}v_{k}^{(ij)}| \geq  c \sqrt{\dfrac{\log p}{ n}}  \right) \leq 2p^{ - c^{2}/ (6e^2 \kappa_{1}^{\prime 2} )}  =2p^{ - c^{2}C_{\text{min}}/ (24e^2 C_{\text{max}} \kappa^4)}.
	\end{equation}
	Since \eqref{ineq:Bernstein} is true for all $i,j$, we have:
	\[
	\P \left(  |\bm{\Omega}_{\bbeta}\bSigma -(\biden-P_C)|_{\infty}  \geq 
	c \sqrt{(\log p) / n} \right) \leq   2p^{ - c^{2}C_{\text{min}}/ (24e^2 C_{\text{max}} \kappa^4)+2} = 2p^{-c_{1}^{''}}.
	\]
	Then by the following inequality:
	\begin{align*}
	&\P \left(  |\bm{\Omega}_{\bbeta}\widehat{\bSigma} -(\biden-P_C)|_{\infty}  \geq  c \sqrt{(\log p) / n} \right) \\
	&   \leq  \P \left(  |\bm{\Omega}_{\bbeta}\bSigma -(\biden-P_C)|_{\infty} + |\bm{\Omega}_{\bbeta}(\bSigma - \widehat{\bSigma})| \geq  c \sqrt{(\log p) / n} \right) \\
	& \leq \P \left(  |\bm{\Omega}_{\bbeta}\bSigma -(\biden-P_C)|_{\infty}  \geq  c \sqrt{(\log p) / n} \right) + \P \left( |\bm{\Omega}_{\bbeta}(\bSigma - \widehat{\bSigma})| \geq  c \sqrt{(\log p) / n}\right)
	\end{align*}
	Notice that:
	\begin{align*}
	 \left|\bm{\Omega}_{\bbeta}(\bSigma - \widehat{\bSigma})\right|_{\infty} & = \frac{1}{n} \left| \sum_{k=1}^{n} \left( \bm{\Omega}_{\bbeta} \left( v(\bbeta, \widetilde{Z}_{k}) - v(\hat{\bbeta}^n, \widetilde{Z}_{k}) \right)  \widetilde{Z}_{k}\widetilde{Z}_{k}^\top \right)  \right|_{\infty}\\
	& \leq \frac{1}{n} \left| \sum_{k=1}^{n} \left( C\|\hat{\bbeta}^{n} - \bbeta\| _1\bm{\Omega}_{\bbeta}  \widetilde{Z}_{k}\widetilde{Z}_{k}^\top \right)  \right|_{\infty}
	\end{align*}
	As 
	\[
	\frac{1}{n} \sum_{k=1}^{n} \left( \bm{\Omega}_{\bbeta}  \widetilde{Z}_{k}\widetilde{Z}_{k}^\top \right) \rightarrow \E \bm{\Omega}_{\bbeta}  \widetilde{Z}_{1}\widetilde{Z}_{1}^\top = \E \bm{\Omega}_{\bbeta}\Theta,
	\]
	together with the result we obtain from theorem \ref{thm: consistency}, 
	\begin{align*}
	\P \left( |\bm{\Omega}_{\bbeta}(\bSigma - \widehat{\bSigma})|_{\infty} \geq  c \sqrt{(\log p) / n}\right) & \leq  \P \left( \frac{1}{n} \left| \sum_{k=1}^{n} \left( C\|\hat{\bbeta}^{n} - \bbeta\| _1\bm{\Omega}_{\bbeta}  \widetilde{Z}_{k}\widetilde{Z}_{k}^\top \right)  \right|_{\infty} \geq  c \sqrt{(\log p) / n}\right) \\
	& \leq 2p^{1 - \hat{c}^{2}/(2K^2)}  = 2p^{-c_{2}^{''}}
	\end{align*}
	where $\hat{c} = \frac{c \phi_{0}}{C|\bm{\Omega}_{\bbeta}\Theta|_{\infty}s(k_0 \tau +1)}$. 
So finally:
	\begin{align*}
	&\P \left(  |\bm{\Omega}_{\bbeta}\widehat{\bSigma} -(\biden-P_C)|_{\infty}  \geq  c \sqrt{(\log p) / n} \right) \\
	& \leq \P \left(  |\bm{\Omega}_{\bbeta}\bSigma -(\biden-P_C)|_{\infty}  \geq  c \sqrt{(\log p) / n} \right) + \P \left( |\bm{\Omega}_{\bbeta}(\bSigma - \widehat{\bSigma})|_{\infty} \geq  c \sqrt{(\log p) / n}\right) \\
	& \leq 2p^{-c_{1}^{''}}+2p^{-c_{2}^{''}}
	\end{align*}

\end{proof}

\subsection*{Proof of Theorem \ref{thm:bound on Delta}}
\begin{proof}
As we obtained in lemma \ref{thm: feasible set}, $\Omega_{\bbeta}$ is in the feasible set with a large probability. That is, event $|M\widehat{\bSigma} -(\biden-P_C)|_{\infty}  \geq  c \sqrt{(\log p) / n}$ happens with large probability. 
    Further more,	
	\begin{align*}
	\P\left(|(\biden-P_C) - M\widehat{\bSigma}^{0}|_{\infty}  \geq  c \sqrt{(\log p) / n}\right) & \leq  \P \left(  |M\widehat{\bSigma} -(\biden-P_C)|_{\infty}  \geq  c \sqrt{(\log p) / n} \right) \\
	 & + \P \left( |M(\widehat{\bSigma}^{0}- \widehat{\bSigma})|_{\infty} \geq  c \sqrt{(\log p) / n}\right).
	\end{align*}
	The bound for the first term on the RHS is the result from lemma \ref{thm: feasible set}. Applying the similar method to the second term, notice that $\|\hat{\bbeta}^{0} - \bbeta\| _1 \leq \|\hat{\bbeta}^{n} - \bbeta\| _1$, hence, $\P ( |M(\widehat{\bSigma}^{0}- \widehat{\bSigma})|_{\infty} \geq \allowbreak c \sqrt{(\log p) / n}) \leq 4p^{-c_{2}^{''}}$.	So, 
	\begin{align*}
	\P\left(|(\biden-P_C) - M\widehat{\bSigma}^{0}|_{\infty}  \geq  c \sqrt{(\log p) / n}\right) \leq 2p^{-c_{1}^{''}}+6p^{-c_{2}^{''}}
	\end{align*}
	Finally, 
	\begin{align*}
	\|\Delta\|_{\infty} & \leq  \sqrt{n} \left|(\biden-P_C) - \tM\widehat{\bSigma}^{0}\right|_{\infty}\|\hat{\bbeta}^n-\bbeta\|_{1} \\
	& =  \sqrt{n} \left|(\biden-P_C)\left((\biden-P_C) - \tM\widehat{\bSigma}^{0}\right)\right|_{\infty}\|\hat{\bbeta}^n-\bbeta\|_{1} \\
	& \leq k_{0} \sqrt{n}|(\biden-P_C) - M\widehat{\bSigma}^{0}|_{\infty}\|\hat{\bbeta}^n-\bbeta\|_{1} 
	\end{align*}
	We have:
	\begin{align*}
	& \P\left(  \|\Delta\|_{\infty} > \dfrac{c\tilde{c}k_{0}(k_0\tau +1)}{\phi_0} \cdot \dfrac{s\log p}{\sqrt{n}}  \right) \\
	& \leq \P \left( \|\hat{\bbeta}^{n} - \bbeta\| _1  \geq \dfrac{s\lambda(k_0+1/\tau)}{\phi_{0}}  = \dfrac{s\tilde{c}(k_{0}\tau+1)\sqrt{(\log p) / n}}{\phi_{0}}\right) \\
	& + \P\left(|(\biden-P_C) - M\widehat{\bSigma}^{0}|_{\infty}  \geq \gamma = c \sqrt{(\log p) / n}\right) \\
	& \leq 2p^{-c'} +2p^{-c_{1}^{''}}+6p^{-c_{2}^{''}}
	\end{align*}
	So we have finished the proof. 
\end{proof}

\bibliographystyle{biom}
\bibliography{glmlassoref}

\label{lastpage}

\end{document}